\documentclass[twocolumn]{aastex631}

\usepackage{CJKutf8}

\usepackage{color}
\usepackage{multirow}
\usepackage{rotating}
\usepackage{longtable}
\usepackage{cancel,soul,ulem,amsmath} 
\usepackage{graphicx}
\usepackage{subfigure}

\def\hi{H{\sc i}}

\def\arcsec{\hbox{$^{\prime\prime}$}}
\def\arcmin{\hbox{$^{\prime}$}}

\def\cm2{cm$^{-2}$}
\def\cc{cm$^{-3}$}
\def\kms{km s$^{-1}$}
\def\s{s$^{-1}$}
\def\nh3{NH$_3$}
\def\n2h{N$_2$H$^+$}
\def\co{$^{12}$CO}
\def\13co{$^{13}$CO}
\def\c18o{C$^{18}$O}
\def\hc3n{HC$_3$N}
\def\h2{H$_2$}
\def\nh{n(H$_2$)}
\def\c2{[C\,{\sc ii}]}

\shorttitle{Physical Properties and \hi-to-\h2\ Transition across Taurus Linear Edge}
\shortauthors{Tang et al.}

\begin{document}

\title{ Physical Properties and \hi-to-\h2\ Transition across Taurus Linear Edge}
\begin{CJK}{UTF8}{gbsn}

\author[0000-0002-2169-0472]{Ningyu Tang}
\affiliation{Department of Physics, Anhui Normal University, Wuhu, Anhui 241002, China}

\author{Feihang Miao}
\affiliation{Department of Physics, Anhui Normal University, Wuhu, Anhui 241002, China}

\author{Gan Luo}
\affiliation{Institut de Radioastronomie Millimetrique, 300 rue de la Piscine, 38400, Saint-Martin d’Hères, France}

\author[0000-0003-3010-7661]{Di Li}
\affiliation{New Cornerstone Science Laboratory, Department of Astronomy, Tsinghua University, Beijing 100084, China}
\affiliation{National Astronomical Observatories, CAS, Beijing 100012, People’s Republic of China}
\affiliation{Zhejiang Lab, Hangzhou, Zhejiang 311121, China}

\author{Junzhi Wang}
\affiliation{Guangxi Key Laboratory for Relativistic Astrophysics, Department of Physics, Guangxi University, Nanning 530004, PR China}

\author{Fujun Du}
\affiliation{Purple Mountain Observatory and Key Laboratory of Radio Astronomy, Chinese Academy of Sciences, Nanjing 210023, China}
\affiliation{School of Astronomy and Space Science, University of Science and Technology of China, Hefei 230026, China}

\author{Donghong Wu}
\affiliation{Department of Physics, Anhui Normal University, Wuhu, Anhui 241002, China}

\author{Shu Liu}
\affiliation{National Astronomical Observatories, CAS, Beijing 100012, People’s Republic of China}

\correspondingauthor{Ningyu Tang}
\email{nytang@ahnu.edu.cn}



\begin{abstract}
Studying the  atomic-to-molecular transition is essential for understanding the evolution of interstellar medium. The linear edge of Taurus molecular cloud, clearly identified in the  $^{13}$CO(1-0) intensity map, serves as an ideal site for investigating this transition.  Utilizing the Arizona Radio Observatory Sub-Millimeter Telescope, we obtained mapping observations of CO(2-1), $^{13}$CO(2-1), and CO(3-2) across this linear edge. The intensity ratio between CO(2-1) and $^{13}$CO(2-1) indicates a lower limit of the $^{12}C/^{13}C$ ratio of $54\pm 17$. Based on multi-transition observations of CO and \13co, we performed  Markov Chain Monte Carlo (MCMC) fit of the physical properties across this edge using non-Local Thermodynamic Equilibrium analysis with the RADEX code, based on the Large velocity Gradient (LVG) assumption. The number density profile exhibits a pronounced jump coinciding with the \h2\ infrared emission peak. 
The cold \hi\ gas within the molecular cloud, manifested as \hi-Narrow Self-Absorption (HINSA) features, is detected along the cloud edge. 
Our quantitative comparison with numerical simulations provides tentative evidence that shocks induced by colliding gas flows may contribute to the atomic-to-molecular phase transition observed along the linear edge.
\end{abstract}



\section{Introduction}
\label{sec:intro}


The transition from  \hi\ gas to  \h2\ gas  is considered  a critical process for understanding the evolution of star formation. Photo-dominated regions  \citep[e.g., PDRs,][]{1999RvMP...71..173H, 2006ARA&A..44..367S}, which are dominated by far-ultraviolet (FUV; 6 eV $\lesssim$ $h$$\nu$ $\lesssim$ 13.6eV) photons, maintain a natural atomic-molecular transition between warm \hi\ envelope and cold \h2 center. The in situ formation of CO has been detected within the warm \hi\ envelope (with an excitation temperature of $\sim 100$ K) located at the boundary of supershells \citep[e.g.,][]{2011ApJ...741...85D, 2015ApJ...799...64D}. In the center of the molecular cloud where FUV photons are well shielded, cold \hi\ gas with a kinetic temperature of $\sim 10$ K can exist  \citep[e.g.,][]{1969ApJ...156..493H, 2003ApJ...585..823L}. The cold \hi\ arises from the dissociation of \h2\ by low-energy cosmic rays (LECRs). Understanding the physical conditions that govern atomic-to-molecular transition inside the molecular cloud is also of paramount scientific interest.

With $^{13}$CO(1-0) data, a clear `linear edge' was observed toward the boundary of Taurus molecular cloud  \citep{2008ApJ...680..428G}. It provides an ideal site for investigating the molecular-to-atomic transition. Toward this linear edge, \citet{2010ApJ...715.1370G} obtained direct \h2\ infrared emission with the $Spitzer\ Space\ Telescope$  and revealed the presence of warm \h2 with low column density, $(1-5)\times 10^{18}$ $cm^{-2}$. Additional heating sources, such as the dissipation of turbulence are necessary to explain this \citep{2010ApJ...715.1370G}. Based on multiple spectral observations (C{\sc ii}, C{\sc i}, CO(1-0) and $^{13}$CO(1-0)),  \citet{2014ApJ...795...26O}  modeled  this region with a cylindrical PDR under low FUV intensity, $\chi \sim 0.05\ G_0$, through combining the chemical PDR model ($Meudon$) and radiative transfer code ($RATRAN$).  
With the density profile from \citet{2014ApJ...795...26O}, \citet{2016ApJ...819...22X} and  \citet{2016ApJ...833...90X} utilized OH and CH emissions to obtain the fraction of CO-dark molecular gas across this edge. However, the above modeling physical conditions depend on the assumption of chemical abundances \citep{2014ApJ...795...26O}.

The combination of multiple-$J$ CO transitions provides a better solution in deriving physical conditions of molecular clouds. For instance, with assumption of optically thin emission and kinetic temperature of 10 K, the critical density  $n\rm_{crit}$ of CO(3-2) is $4\times 10^4$ \cc,  which is much larger than that of CO(1-0) ($\sim 2\times 10^3$ \cc) and CO(2-1) ($\sim 1\times 10^4$ \cc)  \citep{1987ApJS...63..821S}. The critical density of  CO(1-0) and CO(2-1) can be reduced to $\sim 10^2-10^3$ \cc\ if optical depth is included.  Thus the combination of multiple CO transitions is useful  to determine both kinetic temperature and volume density. With the Arizona Radio Observatory Sub-Millimeter Telescope (SMT), CO surveys in multiple low-$J$ transitions are able to derive physical conditions toward nearby molecular clouds (W51, NGC 1333, SH2-235, and Cep B/C)  \citep{2010ApJS..191..232B, 2014ApJS..214....7B, 2016ApJS..226...13B, 2018ApJS..238...20B}. In diffuse region, CO excitation can not reach local thermodynamic equilibrium (LTE) and requires non-LTE analysis. 

In observations, the \hi\ Narrow Self-Absorption (HINSA) feature is considered to arise from cold (T $\sim 10$ K) atomic hydrogen mixed with molecular gas with visual extinction $A_V > 1$ mag  \citep{2003ApJ...585..823L}, where FUV radiation can be well-shielded and ionization process of low-energy cosmic rays dominate. The HINSA feature, with same central velocity as CO emission, appears across this linear edge. Analyzing molecular-to-atomic transition across this edge would benefit in constraining the insitu flux of low-energy cosmic rays. 

In this study, we focus on exploring the  physical properties and the atomic-to-molecular transition  across the Taurus linear edge. We detail the observations and archival data in Section \ref{sec:obs}, while the analysis and results concerning physical properties and HINSA column density are outlined in Section \ref{sec:analysis}. Discussions regarding the results are provided in  Section \ref{sec:discussion}, and a concise summary is presented in Section \ref{sec:summary}.

\section{Observations and Archival Data}
\label{sec:obs}
\subsection{Observation Targets}
\begin{figure*}
\centering
\includegraphics[width=0.9\textwidth]{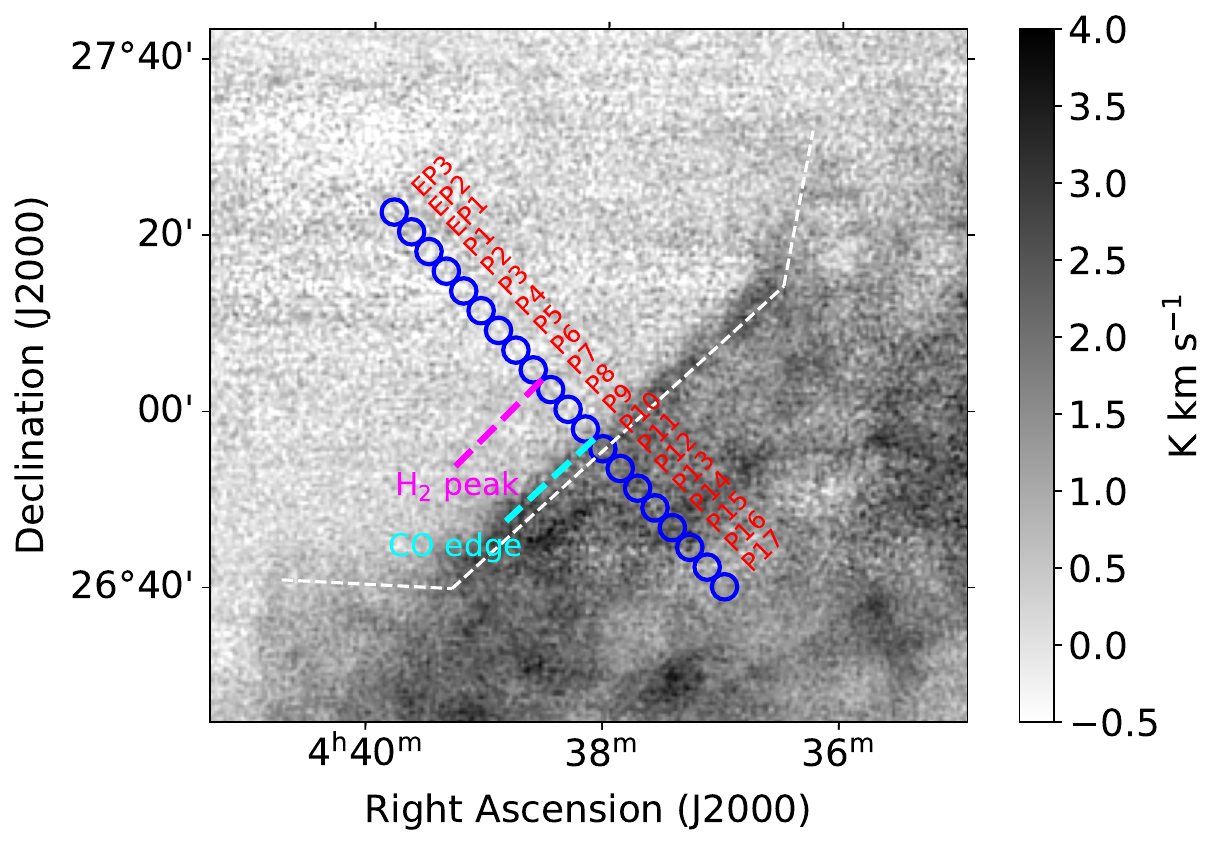}
\caption{Spatial distribution of 20 positions (blue circles) overlaid on $^{13}$CO intensity map across the Taurus linear edge. Blue circles represent observation targets with beam size of 3 arcmin. The peak intensity of the $S(0)$ and $S(1)$ transitions of \h2\ locates between P6 and P7 \citep{2010ApJ...715.1370G}. Apparent CO cloud edge locates between P9 and P10.  The presence of a U-shape is marked with white dashed lines. }
\label{fig:boundpos}
\end{figure*}

As shown in Fig. \ref{fig:boundpos}, we focus on the variation of physical properties across the Taurus linear edge. We select a total of 20 positions, of which 17 positions (P1 to 17) are the same as those in \citet{2016ApJ...819...22X} and  \citet{2016ApJ...833...90X}. In order to explore physical properties in extended regions, we included 3 more positions (EP1, EP2 and EP3). The separation between two nearby positions is 3\arcmin. The intensity peak of \h2\ is located in the middle of P6 and P7 \citep{2010ApJ...715.1370G} while the peak of \13co(1-0) intensity is located between P9 and P10 \citep{2008ApJ...680..428G}.  

\subsection{J=2-1 and J=3-2 Transitions of CO}
\label{sec:co21}

The J=2-1 transitions of CO (centered at 230.538 GHz) and $^{13}$CO (centered at 220.39868420 GHz) were obtained in November 2022 with 1.3 mm receiver of the SMT. The beam size is $\sim$ 33 \arcsec\ for CO(2-1) and $\sim$ 35 \arcsec\ for $^{13}$CO(2-1). We adopted the On the Fly (OTF) mapping mode to cover a map with  size of  5$'$ $\times$ 3$'$ (along R.A and Dec. direction) direction around each position.   Typical system temperature ranged between 200 and 275 K during observations. We choose backend  with 64 MHz bandwidth and channel spacing of 250 kHz corresponding to velocity of $\sim 0.33$ \kms\ at 230 GHz.  

The J=3-2 transitions of CO (centering at 345.796 GHz) were obtained in February and March 2024 with 0.8 mm receiver of the SMT. Due to time constraints, obtained map around each position has a size of 3$'$ $\times$ 3$'$ (along R.A and Dec. direction) except P2, which has a size of 5$'$ $\times$ 3$'$. The beam size is $\sim$ 22\arcsec\ for CO(3-2). The system temperature ranged from 700 to 1500 K depending on elevation angle and weather condition during observations. Selected channel spacing of 250 kHz  provides  velocity  of $\sim 0.22$ \kms\ at 345 GHz.

\begin{figure*}
\centering
\includegraphics[width=0.32\textwidth]{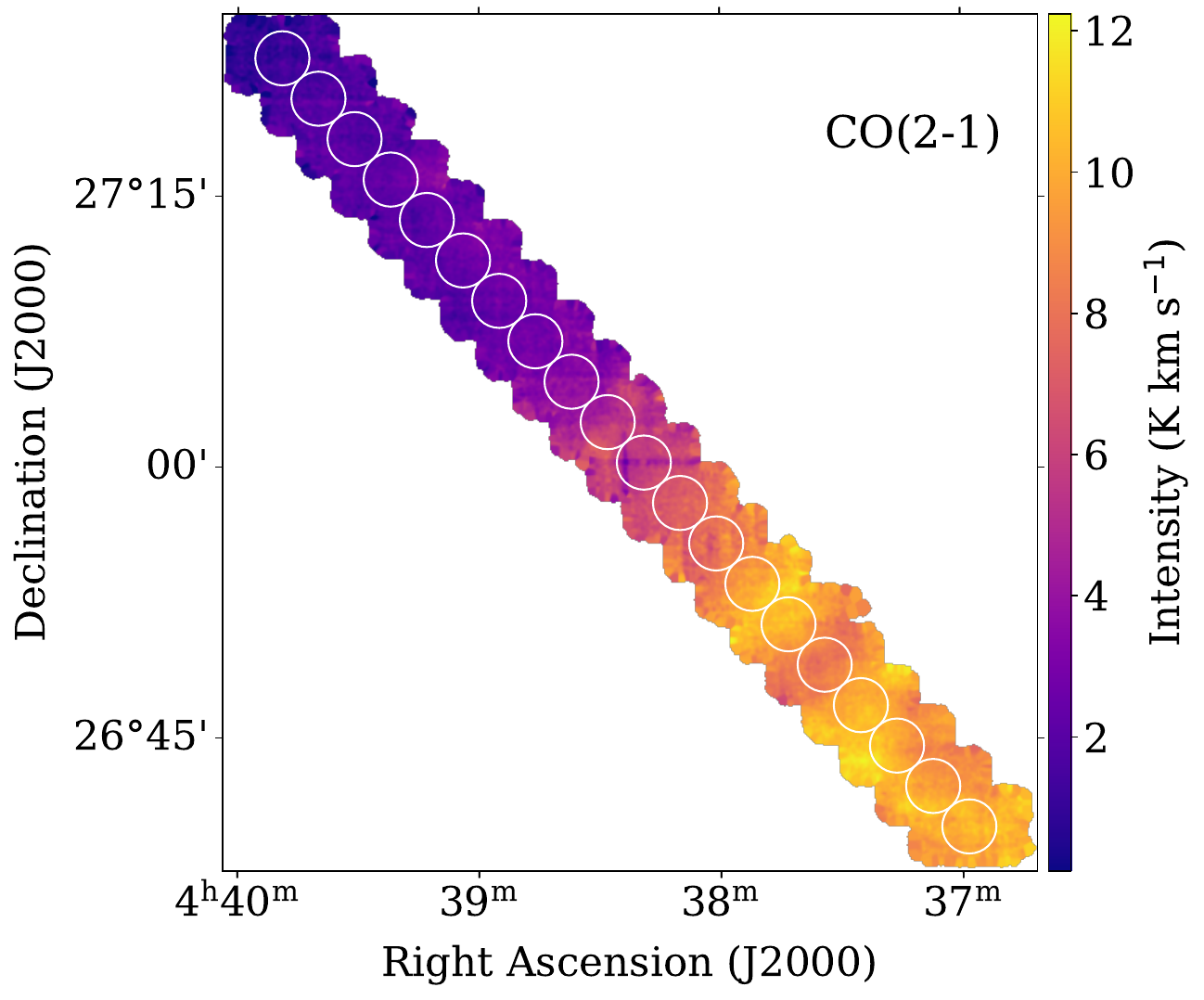}
\includegraphics[width=0.33\textwidth]{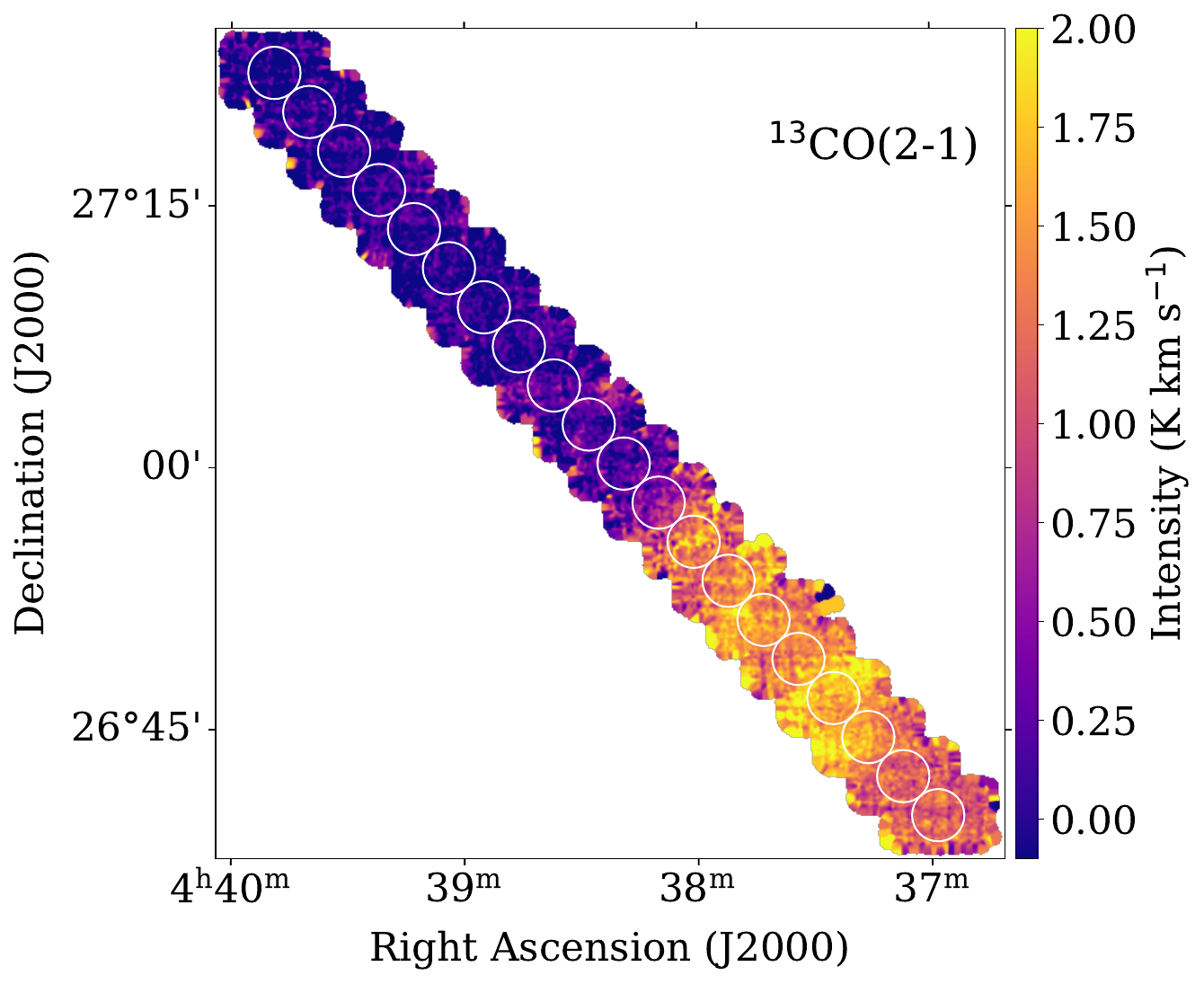}
\includegraphics[width=0.305\textwidth]{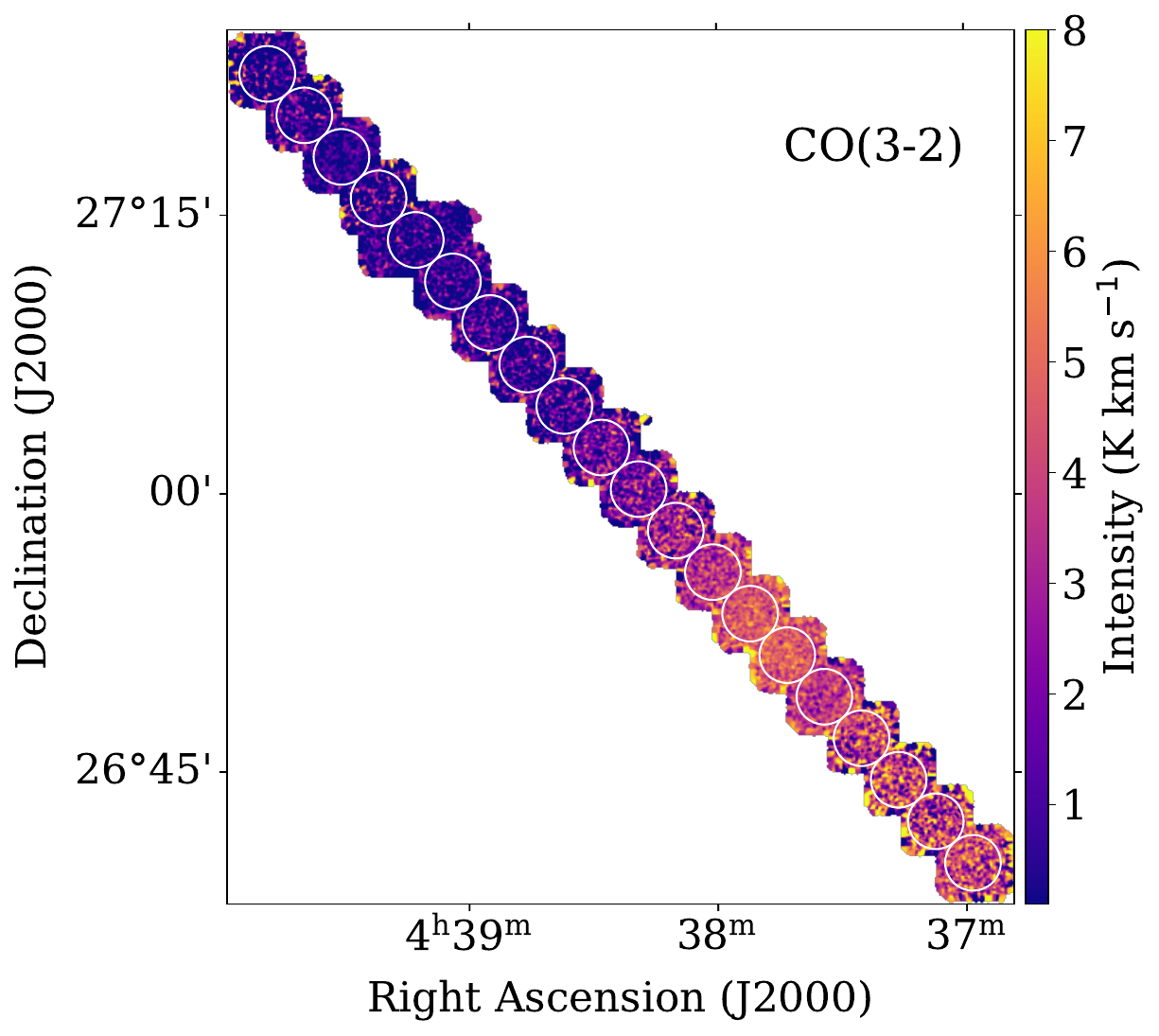}
\caption{Obtained intensity  of CO(2-1), \13co(2-1) and CO(3-2) integrated over $-15$ to 15 \kms\ across the linear edge. The white circles represent 20 positions with beam size of 3\arcmin. }
\label{fig:intg_map}
\end{figure*}

Data reduction was performed with the $CLASS$ program in GILDAS package\footnote{https://www.iram.fr/IRAMFR/GILDAS/}. Bad scans with invalid values were first flagged. All reduced maps were combined together into one map by coordinate. With pixel sampling of 4\arcsec\ , final root-mean-square of the spectrum of each pixel (in $T\rm_A$ unit) is $\sim 0.14$ K per 0.33 \kms\  for CO(2-1) and $^{13}$CO(2-1), while it is $\sim 0.68$ K per 0.22 \kms\ for CO(3-2).  The intensity map of these transition lines integrated over the velocity range $-15$ to 15 \kms\  can be found in Fig. \ref{fig:intg_map}.

The antenna temperature $T\rm^{*}_A$ is converted to main beam brightness temperature $T\rm_{mb}$ using $T\rm_{mb}$ = T$\rm^{\ast}_{A}$/$\eta\rm_{eff}$ , where an efficiency factor $\eta\rm_{eff}$ = 0.70 was adopted for 1.3mm receiver, and $\eta\rm_{eff}$ = 0.62 for the 0.8mm receiver\footnote{https://aro.as.arizona.edu/?q=beam-efficiencies}. The half-power beamwidth of SMT is 34$''$ at 230 GHz, 36$''$ at 220 GHz and  23$''$ at 345 GHz. 

\subsection{ CO(1-0) Data}
\label{sec:co10}

The $^{12}$CO (1-0) (centering at 115.271 GHz)  and $^{13}$CO (1-0) (centering at 110.201 GHz)  observations were taken with the 14 m diameter millimeter-wavelength telescope of Five College Radio Astronomy Observatory (FCRAO) between 2003 and 2005. The observations have full width at half-maximum of 45\arcsec\ for \co\ and 47\arcsec\ for \13co  \citep{2008ApJS..177..341N}. Main beam efficiency of 0.45 and 0.50 were used to convert antenna temperature to main beam brightness temperature of CO J=1-0 and $^{13}$CO J=1-0, respectively  \citep{2008ApJ...680..428G}.  The data (in unit of T$\rm_{mb}$)  have a mean rms  of 0.62 K and 0.25 K per 0.26 and 0.27 km s$^{-1}$ for $^{12}$CO and $^{13}$CO, respectively.

The signal-to-noise ratio of the $^{13}$CO(1-0)  emission from EP3 to P3 is relatively low. To enhance sensitivity at these 6 positions, new $^{13}$CO(1-0) mapping observations were taken with the Delingha 13.7m telescope between January 10th and 12th, 2025, covering an area of 22\arcmin $\times$ 10\arcmin.  Simultaneously, corresponding $^{12}$CO(1-0) mapping data were also obtained. The system temperature was  $\sim$ 150 K for $^{13}$CO(1-0)   and $\sim $ 250 K for $^{12}$CO(1-0) observations.  The  spectral resolution of 61 kHz corresponds to a velocity resolution of  $0.17$ \kms\ at 110.201 GHz.  The sampling size of the map is 24.5\arcsec, which is half  the main beam size of the Delingha 13.7m telescope at 110.201 GHz.  After processing with $GILDAS$ software, rms of the spectrum (in unit of T$\rm_{mb}$) for each pixel reaches  0.12 K for $^{12}$CO  and 0.08 K  for $^{13}$CO, corresponding to velocity resolutions of 0.16 and 0.17 km s$^{-1}$, respectively.  

The peak values of $^{12}$CO (1-0) emissions from EP3 to P3 differ by 5\% to 15\% between FCRAO and Delingha 13.7m observations, attributed to uncertainties in main-beam efficiency. To ensure  data consistency across the 20 positions, the $^{13}$CO(1-0) data from Delingha 13.7m  were adjusted by multiplying with the peak intensity ratio of  $^{12}$CO(1-0) between FCRAO and Delingha 13.7m  observations.

\subsection{Archival \hi\ Data}
\label{sec:hi}

 \hi\ data across this linear edge were extracted from the Data Release 2 of Galactic Arecibo L-Band Feed Array \hi\ (GALFA-\hi) survey, which provides angular resolution of $\sim$ 4$'$ and spectral sensitivity of $\sim$ 150 mK (brightness temperature) per 1 km s$^{-1}$  \citep{2018ApJS..234....2P}.

\subsection{Archival E(B-V) Data}
\label{sec:EBV}

We adopted the three-dimensional (3D) dust reddening E(B-V) map from  \citet{2019ApJ...887...93G}, which is based on Gaia parallaxes and stellar photometry from Pan-STARRS 1 and 2MASS. This 3D E(B-V) map has a spatial resolution of $3.4\arcmin-13.7\arcmin$ and covers 120 distance bins from 63 pc to 63 kpc. The E(B-V) value of each position in this study was adopted as cumulative value with distance of 300 pc, which is far enough to include the Taurus molecular cloud  \citep[$\sim$ 140 pc, ][]{2013A&A...550A..38P}  and sufficient stars for calculation.  The E(B−V) values within a 300 pc truncation  exceed 90\% of those measured along the entire sightline.

\section{Analysis and Results}
\label{sec:analysis}

\subsection{Spectral Properties}
\label{subsec:spec_prop}

To mitigate the impact of varying spatial resolutions in observations, the multi-transitions of CO data were convolved to a beam width of 3\arcmin\ to match the spatial separation of two nearby positions.  

Spectra of CO and \13co\ transitions are shown in Fig \ref{fig:spec}.  We adopted Gaussian  fitting for all spectra.  One Gaussian component was used for positions EP3 to P13, whereas two components were employed for positions P14 to P17.  Fitted parameters are shown in Table \ref{table:ohfitresult}. For the first component, the central velocity of $^{13}$CO transitions  shifts from $6.51$ \kms\ of position P1  to  5.38 \kms\ of P10 and then reaches  $5.95$ \kms\ of position P17.  A significant second component of CO(1-0) and CO(2-1) spectra arises from $\sim 7.7$ \kms of P14 to $\sim 8.6$ \kms\ of P17. In this study, we focus on the physical properties of the first velocity component. Discussion about the second velocity component is  in Section \ref{sec:discussion}.

 \citet{2016ApJ...824..141G}  confirmed similar velocity gradient in the boundary of L1599B and treated this as the rotation of the central portion of the cloud about its long axis. However, the velocity gradient in this linear edge is not monotonic, implying a low possibility of  the cloud rotation.  This velocity gradient is also seen in both OH and CH spectra and is interpreted as the effect of C-shock  \citep{2016ApJ...819...22X, 2016ApJ...833...90X}. 

\begin{figure}
\centering
\includegraphics[width=0.5\textwidth]{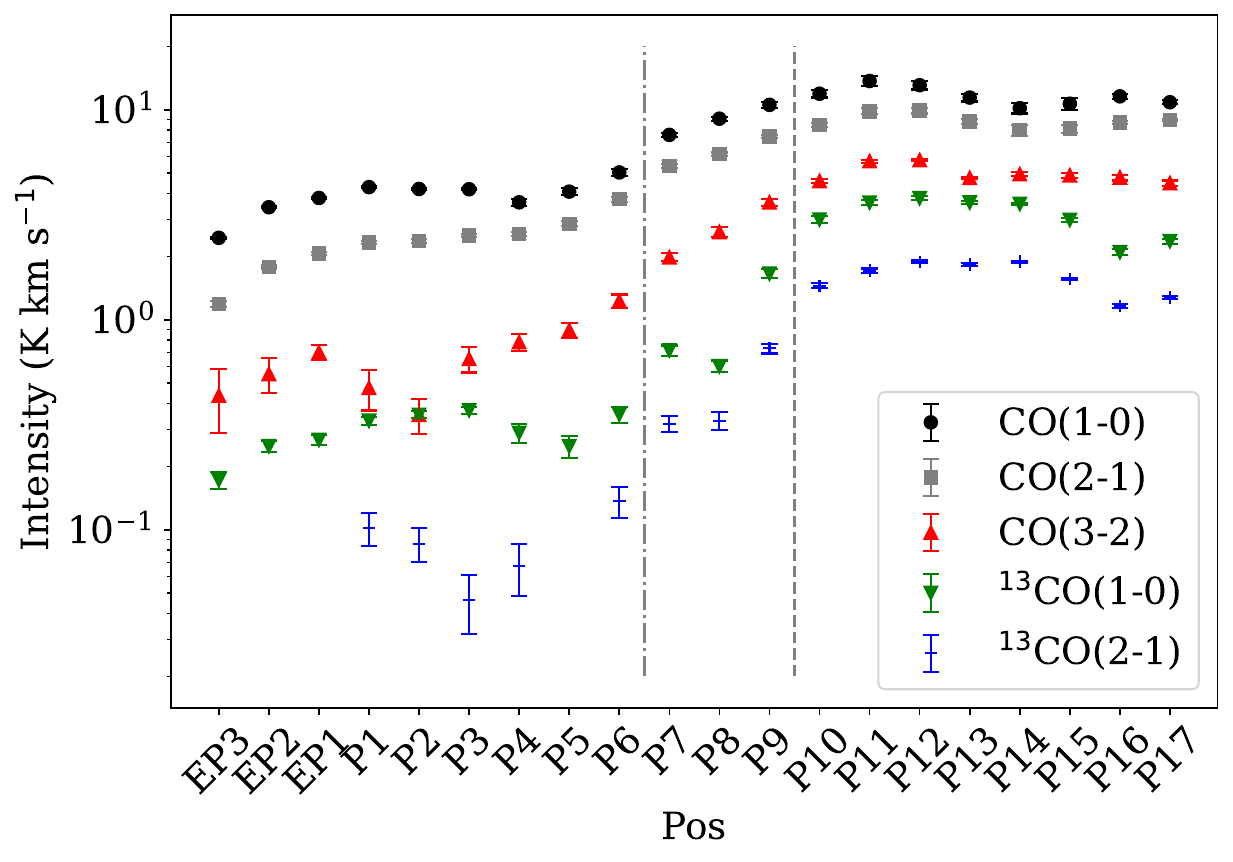}
\caption{ Integrated intensity  of CO/$^{13}$CO(1-0), CO/$^{13}$CO(2-1), CO(3-2) of the first velocity component across Taurus linear edge. The vertical dash-dotted and dashed lines represent the spatial locations of  the \h2\ peak and CO edge shown in Fig. \ref{fig:boundpos}.}
\label{fig:intg_boundary}
\end{figure}

The integrated intensity of multi-J transitions of CO and $^{13}$CO toward the first velocity component is presented in Fig. \ref{fig:intg_boundary}. The intensity values of CO(1-0) and CO(2-1) reaches maximum at P9, the position outside linear edge.   The intensity of other three lines reaches maximum value toward P12, the position inside linear edge.

\subsection{Ratios of Integrated Intensity}
\label{subsec:spec_ratios}

\begin{figure}
\centering
\includegraphics[width=0.5\textwidth]{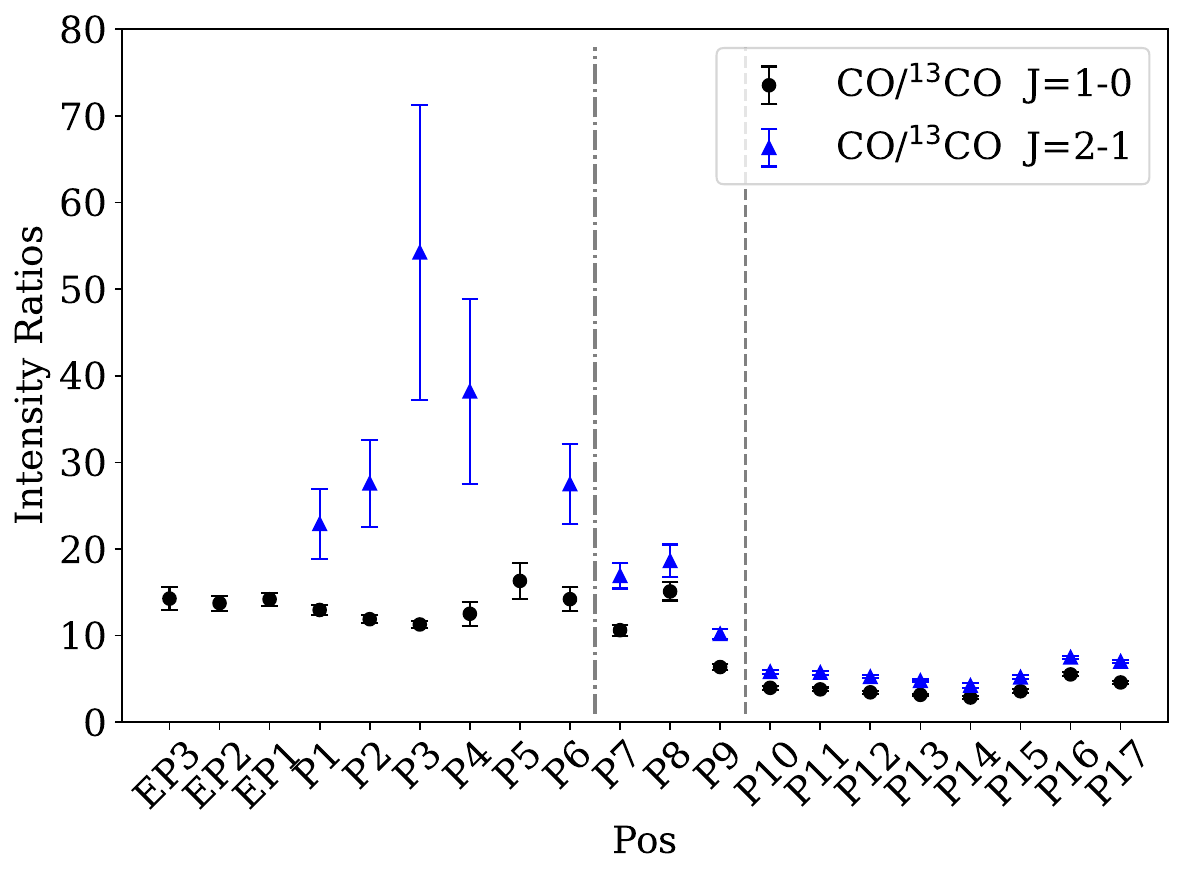}
\caption{Intensity ratio between CO and $^{13}$CO in J=1-0 and J=2-1 transitions of the first velocity component across Taurus linear edge. The vertical dash-dotted and dashed lines represent the spatial locations of  the \h2\ peak and CO edge shown in Fig. \ref{fig:boundpos}.}
\label{fig:ratio_1213}
\end{figure}

We first inspect the intensity ratio of $^{12}$CO and $^{13}$CO in the J=1-0 transition, $R_{1-0}$ and the intensity ratio of $^{12}$CO and $^{13}$CO in the  J=2-1 transition, $R_{2-1}$ toward the first velocity component.  These two ratios are affected by both the $\rm ^{12}C/^{13}C$ value and optical depth.  As shown in Fig. \ref{fig:ratio_1213}, $R_{1-0}$ does not exceed 20 while $R_{2-1}$ reaches a maximum of 54 outside the linear edge. This discrepancy stems from the non-negligible opacity of CO(1-0).  With the assumption of optically thin and same excitation temperature of CO and $^{13}$CO, the $R_{2-1}$ value toward P3 provides an isotopic $\rm ^{12}C/^{13}C$ ratio of $54\pm 17$ in this region. This value represents a lower limit considering opacity effects. 

Our finding of the isotopic $\rm ^{12}C/^{13}C$ ratio is consistent with the indicated low ambient value of 43 across this edge \citep{2014ApJ...795...26O}  and the value of $59-76$ in local ISM  \citep{1998A&A...337..246L, 1999RPPh...62..143W, 2005ApJ...634.1126M, 2007ApJ...667.1002S, 2008A&A...477..865S, 2011ApJ...728...36R}. However, this isotopic value is much lower than that of $\sim 90$ based on $^{13}$CO isotopologs of HCCNC and HNCCC observations toward TMC-1 \citep{2024A&A...682L..13C}.

\begin{figure}
\centering
\includegraphics[width=0.5\textwidth]{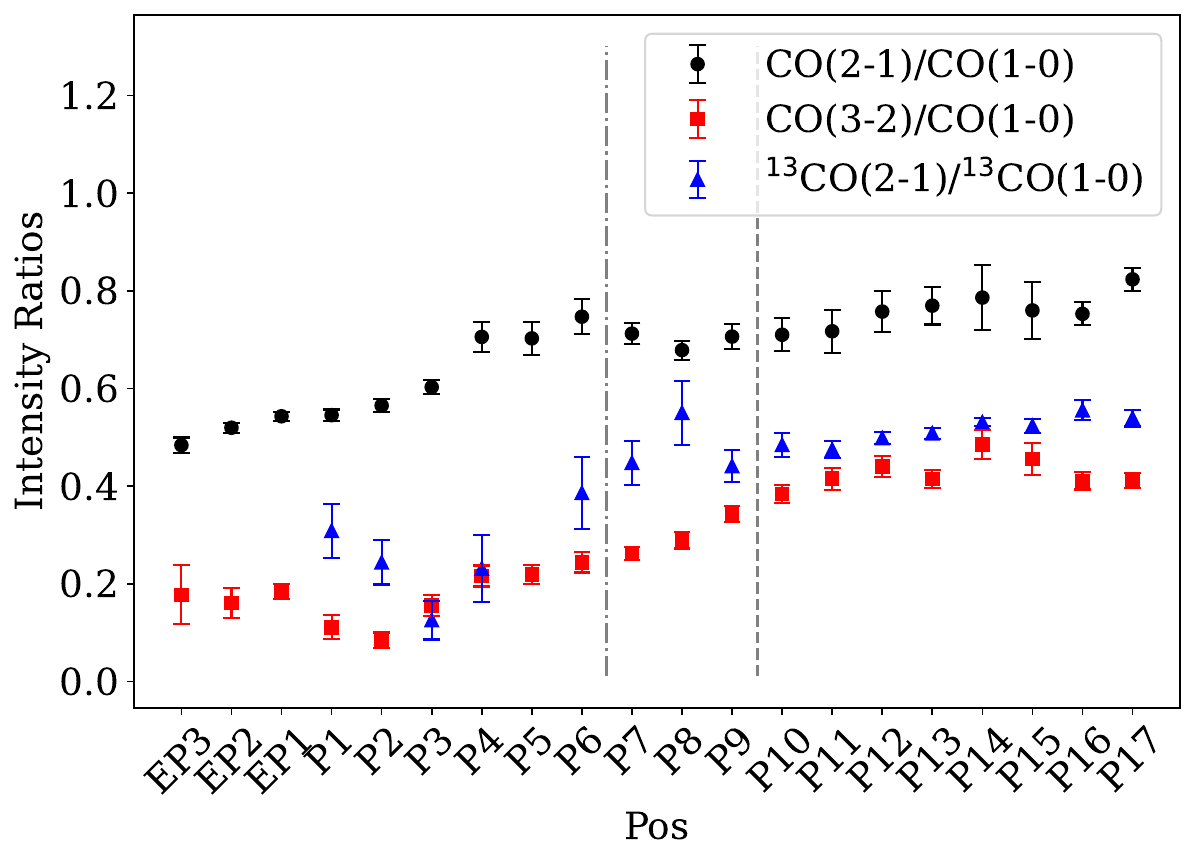}
\caption{The intensity ratio between CO(2-1) and CO(1-0) of the first velocity component across Taurus linear edge. The vertical dash-dotted and dashed lines represent the spatial locations of  the \h2\ peak and CO edge shown in Fig. \ref{fig:boundpos}.}
\label{fig:ratio_21_31}
\end{figure}

Intensity  ratio of CO(2-1)/CO(1-0)  (hereafter, R$^{12}_{21}$) and $^{13}$CO(2-1)/$^{13}$CO(1-0) (hereafter, R$_{21}^{13}$) are  sensitive to kinetic temperature and volume density.  As shown in Fig. \ref{fig:ratio_21_31}, R$^{12}_{21}$ value varies from  0.48  to 0.82 with a median value of 0.70. The R$_{21}^{13}$ value is systematically lower. It varies from 0.13 to 0.56 with median value of 0.48.  

For optically thick transitions,  excitation temperature $T\rm_{ex}$ is related with main beam brightness temperature $T\rm_{mb}$ through \citep[e.g.,][]{1975ampi.proc..373P,1991ApJ...374..540G,2016ApJ...824..141G},

\begin{equation}
T_{ex} =  \frac{h\nu/k}{ln\left(1+ \frac{h\nu/k}{T_{mb}+ B}  \right)} ,
\end{equation}
where $h$,$k$, and $\nu$ are Planck constant, Boltzmann constant and frequency of the transition. $B= \frac{h\nu/k}{e^{h\nu/(kT_{bg})}-1}$, where $T_{bg}$ is taken as the cosmic microwave background of 2.73 K. Derived $T_{ex}$ value for multi-J transitions of CO is shown in Table \ref{table:fitpara}.  The $T_{ex}$ value varies from $\sim 5$ K toward outermost position to maximum value of $\sim$ 10 K.

Since $^{13}$CO(2-1) and $^{13}$CO(1-0) transitions are optically thin, their intensity ratio R$_{21}^{13}$ and the excitation temperature of J=2-1, T$_{ex}$(2-1) are connected by the equation  $R_{21}^{13}= \frac{\int T_{mb}(2-1) dv}{\int T_{mb}(1-0) dv}= 4 e^{-10.6/T_{ex}(2-1)}$ \citep{1975ApJ...196L..39G}. However, this equation adopts $Rayleigh-Jeans$ approximation and ignores the 2.73 K cosmic microwave background. It will underestimate $R_{21}^{13}$ by 20\% for $T_{ex}(2-1)=5$ K and 2\% for $T_{ex}(2-1)=20$ K \citep{1981ApJ...245..495L}. Under optically thin and LTE assumption, we derive an accurate relationship based on equations in  \citet{2015PASP..127..266M}, 

\begin{equation}
\begin{split}
R_{21}^{13}= \frac{\int T_{mb}(2-1) dv}{\int T_{mb}(1-0) dv} &= f\frac{A_{21}g_2\nu_{10}^2}{A_{10}g_1\nu_{21}^2}e^{-\Delta E_{21}/(kT_{ex}(2-1))} \\
& = f\cdot 4e^{-10.6/T_{ex}(2-1)},    
\end{split}
\end{equation}
in which $A\rm_{ul}$ and $\nu\rm_{ul}$ represent the spontaneous emission coefficient  and rest frequency for the transition $u\rightarrow l$ of $^{13}$CO.   $g\rm_{u}=2J_u+1$ is rotational degeneracy of the upper energy level.  The correction factor $f$ is expressed with

\begin{equation}
f = \frac{J_{\nu_{10}}(T_{ex}(2-1))}{J_{\nu_{21}}(T_{ex}(2-1))}\frac{J_{\nu_{21}}(T_{ex}(2-1))- J_{\nu_{21}}(T_{bg})}{J_{\nu_{10}}(T_{ex}(2-1))- J_{\nu_{10}}(T_{bg})}, 
\end{equation}
in which \textit{Rayleigh–Jeans equivalent temperature}, $J_{\nu}(T)= \frac{h\nu/k}{exp(h\nu/kT)-1}$. The $f$ value approaches 1 when $Rayleigh-Jeans$ approximation $\frac{h\nu}{kT_{ex}}\ll 1$ and $T\rm_{bg} \ll T_{ex}$ are satisfied. 

 The $T_{ex}(2-1)$ value of these 20 positions ranges from 2.80 to 4.82 K (see Table \ref{table:fitpara}). The derived value is significantly lower than typical kinetic temperature of 10 K in diffuse molecular regions \citep{2008ApJ...680..428G}, indicating the necessity of non-LTE calculations for $^{13}$CO.  

\subsection{Modeling Multi-transitions of CO with RADEX}
\label{subsec:co-excitaiton}

In order to accurately calculate physical parameters including kinetic temperature, density, and CO column density N(CO), we adopt the RADEX code \citep{2007A&A...468..627V}, which is efficient in deriving physical parameters of clouds under non-LTE and uniform conditions.  Markov Chain Monte Carlo (MCMC) method \citep{2013PASP..125..306F}  is adopted to find optimum solutions to maximize the likelihood function. To minimize the effect of beam filling factor, we use intensity ratios of 4 transition lines including CO(2-1), CO(3-2), $^{13}$CO(1-0) and $^{13}$CO(2-1) with respect to CO(1-0) as input parameters. Since no \13co(2-1) emission is detected toward EP3, EP2, EP1 and P5, we only use intensity ratios of the remaining 3 transition lines as input. The likelihood function is defined with

\begin{equation}
ln\ p = -\frac{1}{2} \sum_k \left [  \frac{(R^k_{obs}- R^k_{model})^2}{{\sigma^k_{obs}}^2}  + ln(2\pi {\sigma^k_{obs}}^2)  \right ] ,
\label{eq:mcmc}
\end{equation}
where the $R^k_{obs}$ and $\sigma^k_{obs}$ are the $k$th intensity ratio and its uncertainty, respectively. $R^k_{model}$ is the modeled intesntiy ratio from RADEX.

A lower value of $^{12}$C/$^{13}$C $=54\pm 17$ is derived in Section \ref{subsec:spec_ratios}. We adopted a fixed $^{12}$C/$^{13}$C ratio of 54 in RADEX calculations, resulting in N(CO)/N(\13co)=54. This would decrease the freedom of parameters and benefit  obtaining better constraints with MCMC fit.

\begin{figure}
\centering
\includegraphics[width=0.5\textwidth]{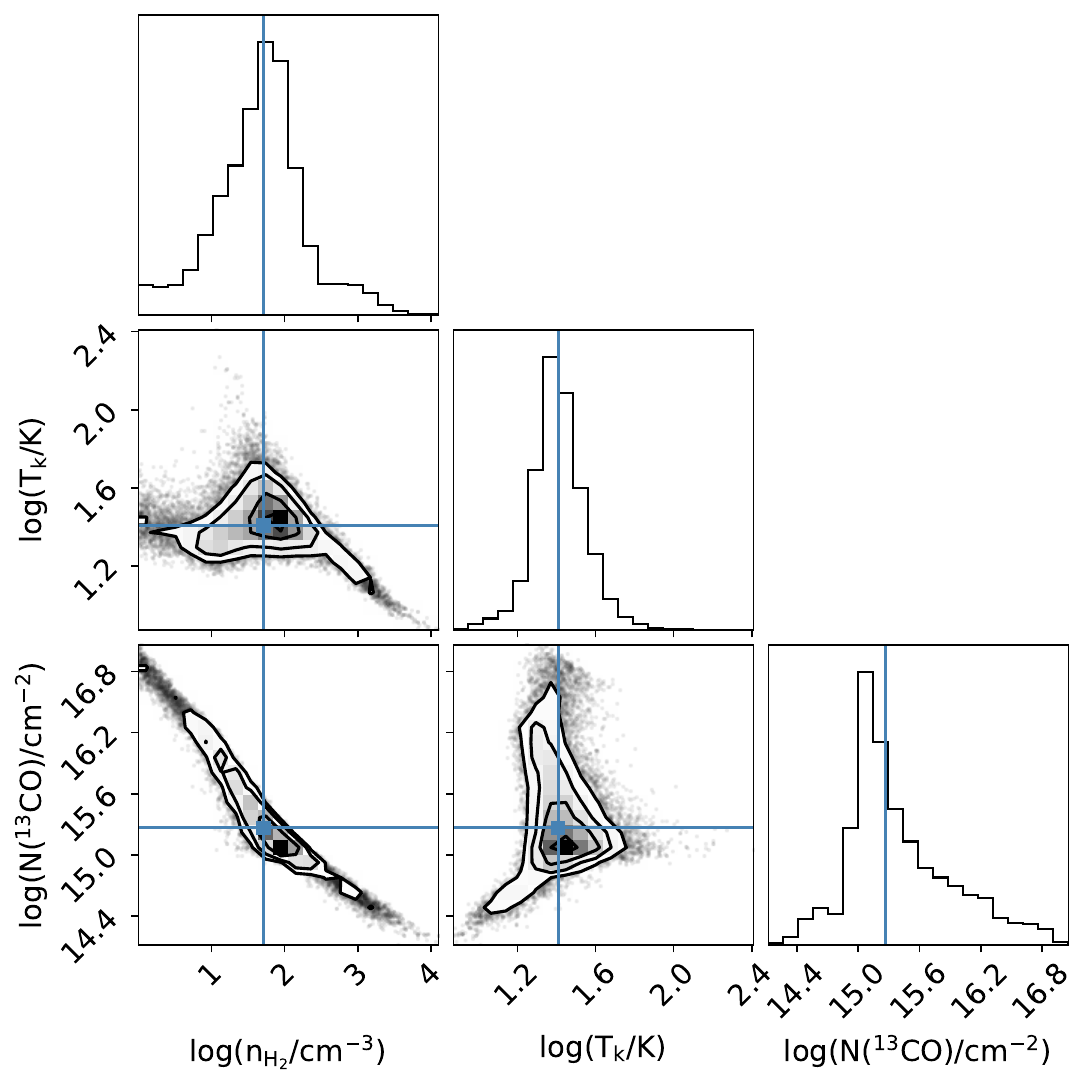}
\caption{Parameter space of MCMC fitting toward P6 under LVG approximation.}
\label{fig:mcmc_fit}
\end{figure}

\begin{figure*}
    \centering
    \includegraphics[width=0.8\linewidth]{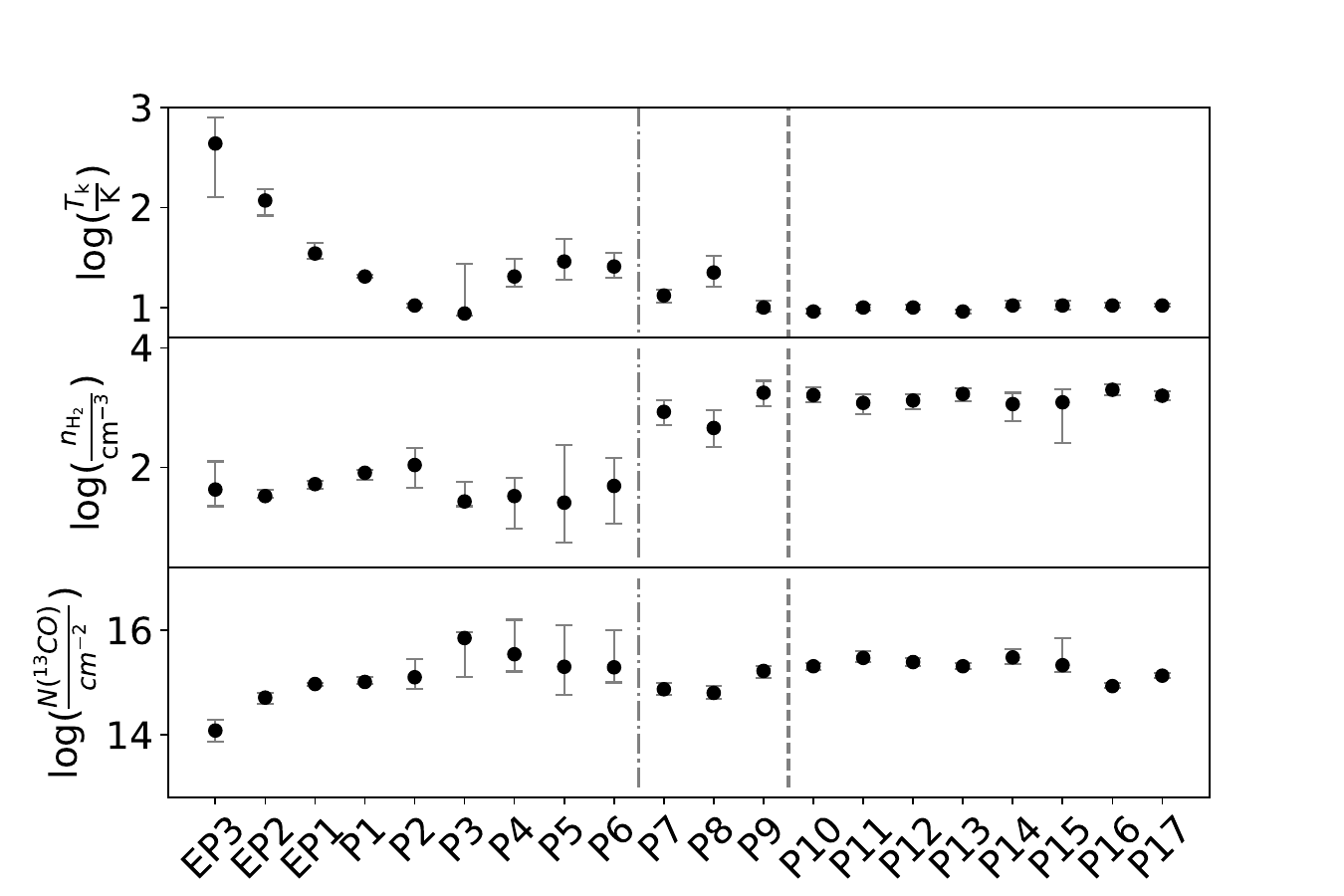}
    \caption{Derived value for physical parameters with MCMC. $Top:$ Kinetic temperature, $T\rm_K$; $Middle:$ Volume density of \h2, $n\rm_{H_2}$; $Bottom:$ Column density of \13co, $N\rm(^{13}CO)$. The vertical dash-dotted and dashed lines represent the spatial locations of  the \h2\ peak and CO edge shown in Fig. \ref{fig:boundpos}.}
    \label{fig:mcmc_paras}
\end{figure*}

With the large velocity gradient approximation \citep[LVG;][]{1957SvA.....1..678S}, RADEX calculations cover the following range of physical parameters:  $5 K \leq  T_k \leq 10^3$ K,  1 \cc $\leq n_{H_2} \leq 10^4$ \cc, and  $10^{13}$ \cm2 $\leq N_{^{13}CO} \leq 10^{18}$ \cm2.  The input line width for RADEX plays a critical role in the modeling process. We utilized the observed line width of \13co(1-0), a transition known to be optically thin. As an example, MCMC fitting toward P6 is shown in Fig. \ref{fig:mcmc_fit}. The value and its uncertainty of each physical parameter are derived by adopting 50 and [16, 84] percentile of MCMC samples. Derived physical parameters toward 20 positions are shown in Table \ref{table:fitpara} and Fig. \ref{fig:mcmc_paras}. 

The MCMC fits converge for three physical parameters toward all 20 positions. The T$\rm_k$ value ranges from 8.71 to 436 K. It decreases from EP3 to P2 and then increases to a peak value of $\sim 28.84$ K toward P5.  The $n\rm_{H_2}$ value varies from 26.9 to $2.0\times 10^3$ \cc. An obvious jump of $n\rm_{H_2}$ value is found from P6 to P7, corresponding to the position with peak \h2\ intensity.  The N($^{13}$CO) value spans from $1.20\times 10^{14}$ to $3.02\times 10^{15}$ \cm2. 
 
A uniform density  was adopted in RADEX calculations. However, the clumpy structure of the molecular gas \citep[e.g.,][]{2008A&A...486L..43H, 2020A&A...633A..51Z} will make the uniform assumption invalid.  With CH observations,  \citet{2016ApJ...833...90X}  found the presence of C-shock in this region, which is not included in the modeling calculations yet.   All of the above discrepancy will introduce uncertainties into our results.  

\begin{longrotatetable}
\begin{deluxetable*}{llllllllllllllllllll}
\tablecaption{Gaussian fitting parameters of CO data. \label{table:ohfitresult}}
\tablewidth{750pt}
\tabletypesize{\tiny}
\tablehead{
\colhead{Pos}  & \multicolumn{4}{c}{CO(1-0)} &  \multicolumn{4}{c}{$^{13}$CO(1-0)} & \multicolumn{4}{c}{CO(2-1)} & \multicolumn{4}{c}{$^{13}$CO(2-1)} & \multicolumn{3}{c}{CO(3-2)}\\ 
\cline{2-4}
\cline{6-8}
\cline{10-12}
\cline{14-16}
\cline{18-20}
\colhead{} &
\colhead{$T\rm_{peak}$} & \colhead{$V\rm_{cen}$} & \colhead{$\Delta V$} &\colhead{} & 
\colhead{$T\rm_{peak}$} & \colhead{$V\rm_{cen}$} & \colhead{$\Delta V$} & \colhead{} &
\colhead{$T\rm_{peak}$} & \colhead{$V\rm_{cen}$} & \colhead{$\Delta V$} &\colhead{} &
\colhead{$T\rm_{peak}$} & \colhead{$V\rm_{cen}$} & \colhead{$\Delta V$} &\colhead{} &
\colhead{$T\rm_{peak}$} & \colhead{$V\rm_{cen}$} & \colhead{$\Delta V$}\\
\colhead{} & 
\colhead{K} & \colhead{\kms} & \colhead{\kms} &\colhead{} & 
\colhead{K} & \colhead{\kms} & \colhead{\kms} &\colhead{} &
\colhead{K} & \colhead{\kms} & \colhead{\kms} &\colhead{} & 
\colhead{K} &\colhead{\kms} & \colhead{\kms} & \colhead{}  &
\colhead{K} &\colhead{\kms} & \colhead{\kms} 
} 

\startdata
EP3 &  2.12(2) &  6.42(1) &  1.08(1) & &  0.11(1) &  6.45(6) &  1.51(6) & & 1.22(3) &  6.39(1) & 0.90(1) & &--- & --- & --- & &  0.40(12) & 6.59(15) &  1.01(15) \\ 
EP2 &  3.25(3) &  6.51(0) &  0.98(0) & &  0.18(1) &  6.51(3) &  1.28(4) & & 1.90(3) &  6.50(1) & 0.88(1) & &--- & --- & --- & &  0.82(14) & 6.56(5) &  0.62(5) \\ 
EP1 &  3.64(4) &  6.61(0) &  0.97(1) & &  0.23(1) &  6.63(2) &  1.07(3) & & 2.14(3) &  6.61(1) & 0.90(1) & &--- & --- & --- & &  0.97(7) & 6.65(2) &  0.67(3) \\ 
P1 &  3.49(4) &  6.70(1) &  1.14(1) & &  0.27(1) &  6.70(2) &  1.12(2) & & 2.04(3) &  6.73(1) & 1.07(1) & & 0.12(2) &  6.96(6) &  0.78(6) & & 0.85(17) & 6.83(5) &  0.52(5) \\ 
P2 &  2.95(3) &  6.70(1) &  1.32(1) & &  0.27(1) &  6.67(2) &  1.24(2) & & 1.79(3) &  6.76(1) & 1.23(1) & & 0.11(2) &  6.91(6) &  0.73(6) & & 0.53(9) & 6.73(5) &  0.62(5) \\ 
P3 &  2.79(3) &  6.72(1) &  1.40(1) & &  0.27(1) &  6.65(2) &  1.29(2) & & 1.82(3) &  6.77(1) & 1.28(1) & & 0.07(2) &  6.86(9) &  0.62(8) & & 0.42(5) & 6.71(9) &  1.44(9) \\ 
P4 &  2.58(8) &  6.61(2) &  1.31(2) & &  0.30(3) &  6.52(4) &  0.90(4) & & 1.75(4) &  6.71(1) & 1.35(1) & & 0.07(2) &  6.58(11) &  0.92(11) & & 0.75(6) & 6.73(4) &  0.97(4) \\ 
P5 &  2.45(9) &  6.54(3) &  1.54(3) & &  0.34(4) &  6.62(4) &  0.69(4) & & 1.71(4) &  6.63(2) & 1.55(2) & &--- & --- & --- & &  0.67(5) & 6.65(5) &  1.25(5) \\ 
P6 &  2.50(9) &  6.37(3) &  1.87(3) & &  0.45(4) &  6.65(3) &  0.74(3) & & 1.79(4) &  6.48(2) & 1.95(2) & & 0.13(2) &  6.76(7) &  1.00(7) & & 0.87(6) & 6.63(4) &  1.31(4) \\ 
P7 &  3.39(7) &  6.26(2) &  2.08(2) & &  0.74(4) &  6.54(2) &  0.90(2) & & 2.30(4) &  6.35(2) & 2.18(2) & & 0.25(2) &  6.76(4) &  1.19(4) & & 1.39(6) & 6.51(3) &  1.33(3) \\ 
P8 &  3.24(7) &  6.01(3) &  2.60(3) & &  0.65(4) &  6.59(2) &  0.85(3) & & 2.25(4) &  6.12(2) & 2.55(2) & & 0.20(2) &  6.62(7) &  1.52(7) & & 1.10(5) & 6.07(5) &  2.20(5) \\ 
P9 &  3.87(10) &  5.78(3) &  2.54(4) & &  0.87(4) &  5.83(4) &  1.76(4) & & 2.74(5) &  5.89(2) & 2.52(2) & & 0.40(2) &  5.87(4) &  1.72(4) & & 1.58(5) & 5.72(3) &  2.12(3) \\ 
P10 &  4.47(16) &  5.66(4) &  2.48(5) & &  2.49(9) &  5.38(2) &  1.12(2) & & 3.21(7) &  5.76(3) & 2.45(3) & & 1.23(3) &  5.53(1) &  1.10(1) & & 2.24(4) & 5.68(2) &  1.90(2) \\ 
P11 &  4.85(22) &  5.62(6) &  2.63(6) & &  2.59(7) &  5.54(2) &  1.30(2) & & 3.71(11) &  5.64(4) & 2.46(4) & & 1.41(3) &  5.65(1) &  1.13(1) & & 3.06(5) & 5.56(1) &  1.73(1) \\ 
P12 &  4.56(19) &  5.93(5) &  2.67(6) & &  3.67(7) &  5.69(1) &  0.96(1) & & 3.77(11) &  5.88(3) & 2.45(4) & & 1.97(3) &  5.76(1) &  0.89(1) & & 3.37(4) & 5.67(1) &  1.59(1) \\ 
P13 &  4.55(17) &  6.15(4) &  2.34(4) & &  3.90(6) &  5.82(1) &  0.86(1) & & 3.56(9) &  6.14(3) & 2.30(3) & & 2.03(2) &  5.90(0) &  0.84(1) & & 3.31(4) & 5.86(1) &  1.33(1) \\ 
P14 &  5.92(31) &  5.74(5) &  1.60(3) & &  3.94(4) &  5.86(0) &  0.84(0) & & 4.49(27) &  5.79(2) & 1.66(3) & & 2.05(2) &  5.92(0) &  0.86(0) & & 3.75(8) & 5.86(1) &  1.22(1) \\ 
P14 &  1.46(14) &  7.70(29) &  2.35(22) & & --- & --- & --- & &  0.90(8) &  7.71(39) & 3.12(26) & &--- & --- & --- & & --- & --- & ---  \\ 
P15 &  5.67(36) &  5.67(4) &  1.75(4) & &  3.47(7) &  5.88(1) &  0.80(1) & & 4.39(17) &  5.75(2) & 1.72(2) & & 1.65(2) &  5.95(1) &  0.88(1) & & 3.99(9) & 5.89(1) &  1.13(1) \\ 
P15 &  1.24(12) &  7.92(41) &  3.12(32) & & --- & --- & --- & &  0.85(6) &  7.92(28) & 3.09(21) & &--- & --- & --- & & --- & --- & ---  \\ 
P16 &  5.54(13) &  5.68(3) &  1.94(3) & &  2.90(8) &  5.88(1) &  0.67(1) & & 4.20(5) &  5.77(2) & 1.93(1) & & 1.29(2) &  6.03(1) &  0.84(1) & & 3.21(9) & 5.84(2) &  1.38(2) \\ 
P16 &  1.22(12) &  8.54(14) &  2.21(15) & & --- & --- & --- & &  0.80(5) &  8.57(9) & 2.28(9) & &--- & --- & --- & & --- & --- & ---  \\ 
P17 &  5.83(12) &  5.80(2) &  1.73(2) & &  2.79(7) &  5.95(1) &  0.79(1) & & 4.58(6) &  5.92(1) & 1.81(1) & & 1.39(2) &  6.11(1) &  0.85(1) & & 3.17(8) & 5.89(2) &  1.31(2) \\ 
P17 &  1.21(11) &  8.61(10) &  2.05(11) & & --- & --- & --- & &  0.86(6) &  8.76(7) & 1.83(7) & &--- & --- & --- & & --- & --- & ---  \\ 
\enddata
\end{deluxetable*}
\end{longrotatetable}

\begin{figure*}
\centering
\includegraphics[width=0.95\textwidth]{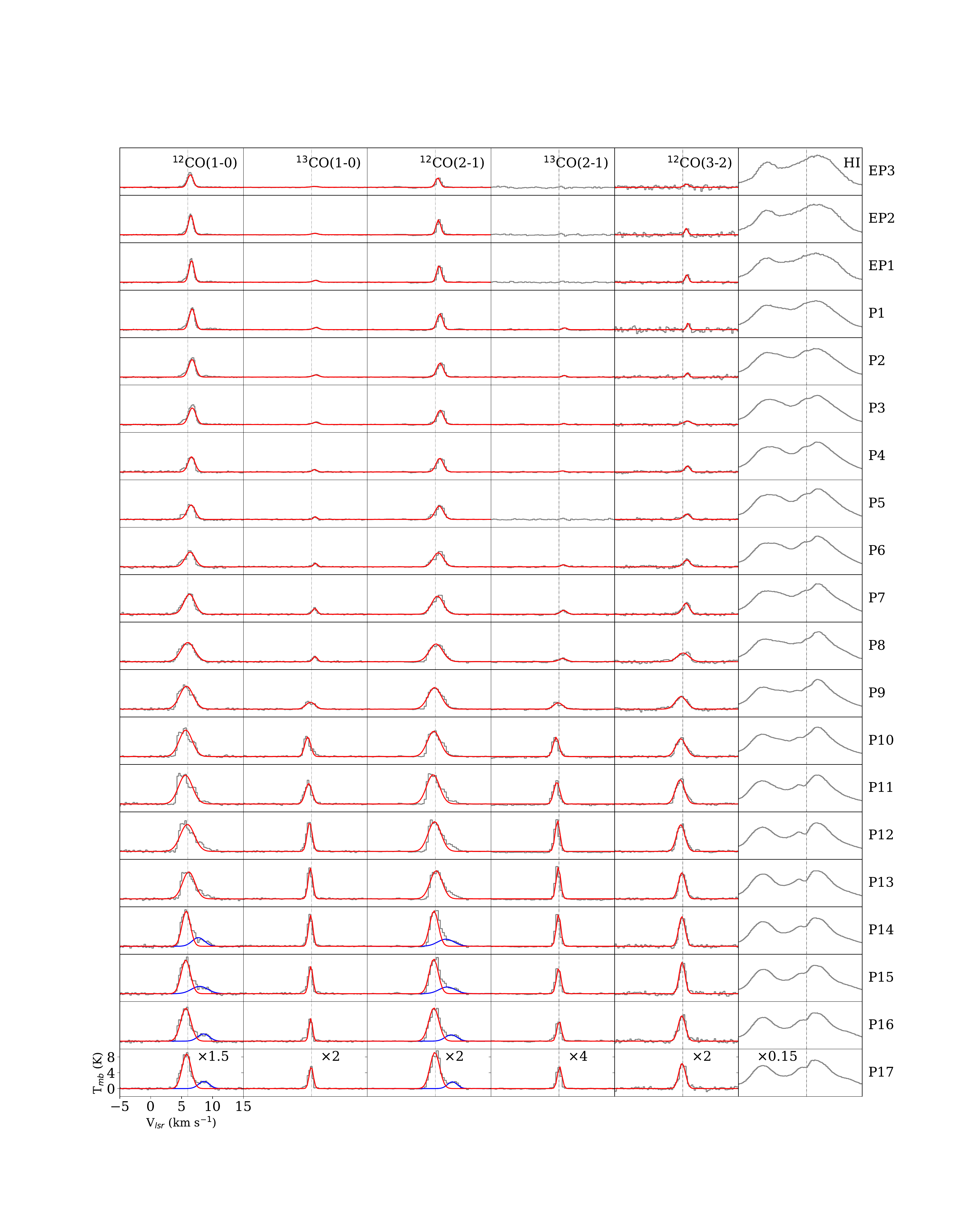}
\caption{Spectra of CO/$^{13}$CO(1-0), CO/$^{13}$CO(2-1), CO(3-2) and \hi\ of the first velocity component across the linear edge. All the spectra were convolved to 3$'$ at each position. The first velocity component of CO is plotted with red Gaussian profile while the second is plotted with blue Gaussian profile. The vertical  dash-dot line represent the position of 6 \kms.  }
\label{fig:spec}
\end{figure*}

\subsection{HINSA Column Density and Abundance}
\label{subsec:hinsa_col}

As seen in Fig. \ref{fig:spec}, an obvious HINSA feature with a central velocity of $\sim 6$ \kms\ can be found from P4 to P17. We follow the HINSA analysis in \citep{2003ApJ...585..823L}  and  \citep{2020RAA....20...77T}. The observed \hi\ brightness temperature can be derived through
\begin{equation}
    T_{ab}(v)=[pT_{HI}(v)+(T_c-T_{ex})(1-\tau_f)](1-e^{-\tau}),
\end{equation}
where $T_c$ and $T_x$ are the background continuum temperature and the excitation temperature, respectively. The value for $T_c$ is derived from the CHIPASS continuum survey at 1.4 GHz  \citep{2014PASA...31....7C}. $T_x$ is assumed to be equal to $T_K$, which is reasonable for \hi\ in the center of molecular clouds, where non-thermal processes are  weak \citep{2017ApJ...843..149S}. The term $\tau_f$ represents the optical depth of the foreground \hi\ gas.  The parameter $p$ is defined as the fraction of $\tau_f$  relative to the total \hi\ optical depth along the line of sight. Given the Taurus cloud's high Galactic latitude, the parameter $p$ can be estimated using the vertical \hi\ distribution of the Milky Way. We adopt the formula for $p$ from \citet{2003ApJ...585..823L}:

\begin{equation}
 p=\rm erfc\left[\frac{\sqrt{4ln(2)} D sin(|b|)}{z}\right],
\end{equation}
where $D$ is the distance to the Taurus cloud, $b$ is the Galactic latitude, $z$ is the FWHM vertical extent of the Galactic \hi\ disk and erfc(x) is the complementary error function, defined as:

\begin{equation}
erfc(x) = 1- (2/\pi^{1/2})\int_0^x e^{-t^2}dt .
\end{equation}

\begin{figure}
\centering
\includegraphics[width=0.5\textwidth]{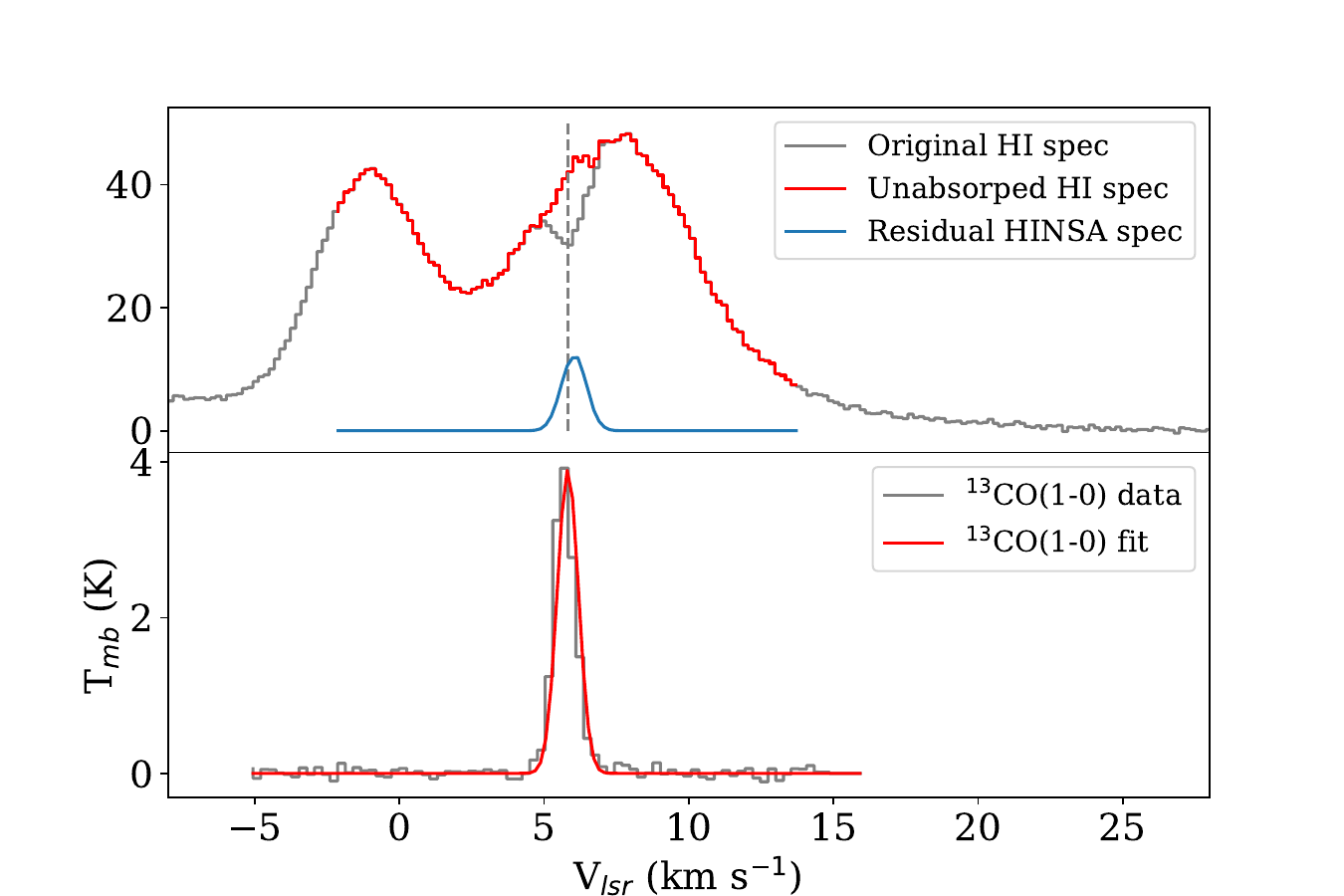}
\caption{HINSA fitting of \hi\ spectrum toward P13.}
\label{fig:hinsafit}
\end{figure}

For $b=-13.0$, the value of $p$ decreases from 0.88 to 0.79 as $D$ varies from 100 pc to 180 pc. In this study, we adopt $p=0.83$, corresponding to $D=140$ pc \citep{2013A&A...550A..38P}. Due to the difficulty in determining an accurate value for $\tau_f$, we assume $\tau_f=0.1$, which may introduce an uncertainty of approximately 4\% in the typical $T\rm_{HI}$ value \citep{2003ApJ...585..823L}. The same values of $p$ and $\tau_f$  are applied to all positions, ensuring that they do not affect the relative HINSA column density across the linear edge. Fig. \ref{fig:hinsafit} shows the example of HINSA fitting toward P6.

After determining optical depth $\tau$, excitation temperature $T_{ex}$ and FWHM of the HINSA emission $\Delta V$, HINSA column density is calculated using the following equation \citep{2003ApJ...585..823L}:  
\begin{equation}
\rm N(HINSA) = 1.9\times 10^{18}\tau T_{ex}\Delta V \rm\ cm^{-2}. 
\end{equation}

The $T\rm_k$ values are adopted from Table \ref{table:fitpara}. The derived HINSA column density of 14 positions is presented in Table \ref{table:fitpara}. The HINSA column density, N(HINSA), ranges from $(5.17 \pm 4.32) \times 10^{18}$ \cm2  at P4 to $(2.65\pm 1.18)\times 10^{19}$ \cm2 at P5. Kinetic temperature significantly influences the calculation of N(HINSA). For instance, a kinetic temperature value of 29 K yields N(HINSA) of $2.7\times 10^{19}$ \cm2 at P5, whereas this value decreases to $3.6\times 10^{18}$ \cm2 if $T\rm_k= 10$ K at the same position. 

\begin{deluxetable*}{lccccccccc}
\tablecaption{Derived physical parameters of the first velocity component. The parameter $T\rm_{ex}^{CO(1-0)}$, $T\rm_{ex}^{CO(2-1)}$, and $T\rm_{ex}^{CO(3-2)}$ represent excitation temperature derived from the optically thick transitions of CO.  $T\rm_{ex}(2-1)$ represents excitation temperature of $J=2-1$ from optically thin transitions of  $^{13}$CO.  T$\rm_k$, n$\rm_{H_2}$,  N(\13co) and N(HINSA) represent fitted kinetic temperature, volume density of \h2, column density of \13co and column density of HINSA, respectively.
\label{table:fitpara}}
\tablehead{
\colhead{Pos} & \colhead{E(B-V)} &  \colhead{$T\rm_{ex}^{CO(1-0)}$ }&  \colhead{$T\rm_{ex}^{CO(2-1)}$} &  \colhead{$T\rm_{ex}^{CO(3-2)}$}   & \colhead{$T\rm_{ex}(2-1)$}     & \colhead{$T\rm{_k}$}  & \colhead{log($\frac{n\rm_{H_2}}{cm^{-3}}$)} & \colhead{log($\rm \frac{N(^{13}CO)}{cm^{-2}})$} & \colhead{N(HINSA)}\\
\colhead{} & \colhead{(mag)} & \colhead{(K)} & \colhead{(K)} & \colhead{(K)} & \colhead{(K)}  & \colhead{(K)} & \colhead{} & \colhead{ } & \colhead{(10$^{18} \rm cm^{-2} $)} 
}
\startdata
EP3 & 0.33 & 5.25 (3)  & 5.09  (5)  &  4.54  (34) & --- & 437$^{+358}_{-311}$ &  1.63$^{+ 0.280}_{- 0.470}$ & 14.1$^{+ 0.2}_{- 0.2}$  & --- \\
EP2 & 0.40  & 6.47  (3) &  6.02  (4) & 5.51  (28) & --- & 117$^{+33.9}_{-34.3}$ &  1.52$^{+ 0.030}_{- 0.110}$ & 14.7$^{+ 0.100}_{- 0.100}$ & --- \\
EP1 & 0.38  &6.88  (4) & 6.33  (3) &  5.81  (14)  & --- & 34.7$^{+ 10.0}_{- 3.77}$ &  1.72$^{+ 0.080}_{- 0.050}$ & 15.0$^{+ 0.030}_{- 0.020}$ & ---\\
P1  & 0.35  &  6.72  (4) & 6.20  ( 4) &  5.58 (34)  & 3.65 (27) &  20.4$^{+ 0.960}_{- 0.460}$ &  1.91$^{+ 0.120}_{- 0.050}$ & 15.0$^{+ 0.040}_{- 0.090}$ & ---\\
P2  & 0.42  & 6.15  (3 )& 5.88  (4) &  4.86  (22)  & 3.31 (24) & 10.5$^{+ 0.490}_{- 0.470}$ &  2.04$^{+ 0.380}_{- 0.280}$ & 15.1$^{+ 0.230}_{- 0.350}$  & ---\\
P3  & 0.44  & 5.97  (3) & 5.92 (5) &  4.59  (14) & 2.80 (26) & 8.71$^{+18.8}_{- 0.390}$ &  1.43$^{+ 0.080}_{- 0.330}$ & 15.9$^{+ 0.740}_{- 0.110}$ & --- \\
P4  & 0.49  &  5.75  (9) & 5.83  (5) &  5.36  (13)  & 3.24  (36) & 20.4$^{+10.5}_{- 4.20}$ &  1.52$^{+ 0.550}_{- 0.300}$ & 15.5$^{+ 0.330}_{- 0.660}$  &  5.17$\pm$4.32\\
P5  & 0.39  & 5.61  (10) & 5.77  (6) &  5.18  (11)  & --- & 28.8$^{+20.1}_{- 9.79}$ &  1.41$^{+ 0.67}_{- 0.97}$ & 15.3$^{+ 0.540}_{- 0.800}$ & 26.5$\pm$11.8\\
P6  & 0.31  &  5.66  (10) & 5.88  (6) &  5.60  (12)  & 4.03 (35) & 25.7$^{+ 9.78}_{- 5.75}$ &  1.69$^{+ 0.630}_{- 0.470}$ & 15.3$^{+ 0.290}_{- 0.710}$ & 11.5$\pm$6.87 \\
P7  & 0.46  &  6.62  (8) & 6.54  (5) &  6.54  (9) & 4.32 (21) & 13.2$^{+ 1.95}_{- 1.96}$ &  2.93$^{+ 0.220}_{- 0.200}$ & 14.9$^{+ 0.110}_{- 0.120}$ & 6.49$\pm$3.00 \\
P8  & 0.53  &  6.46  (7) & 6.47  (5) &  6.05  (9)  &4.80 (30) &  22.4$^{+10.7}_{- 6.17}$ &  2.66$^{+ 0.320}_{- 0.300}$ & 14.8$^{+ 0.120}_{- 0.140}$  & 17.7$\pm$7.18 \\
P9  & 0.67  &  7.12  (11) & 7.09 (6) &  6.86  (8)  &4.29 (15) &  10.0$^{+ 1.75}_{- 0.880}$ &  3.25$^{+ 0.230}_{- 0.200}$ & 15.2$^{+ 0.130}_{- 0.090}$  & 11.6$\pm$3.69 \\
P10 & 0.74  &  7.75 (17) & 7.65  (9) &  7.85  (6)  &4.49 (11) &  9.12$^{+ 0.650}_{- 0.410}$ &  3.21$^{+ 0.120}_{- 0.130}$ & 15.3$^{+ 0.080}_{- 0.070}$ & 9.57$\pm$3.40 \\
P11 & 0.98  & 8.14  (23) & 8.24  (13) &  8.97  (6)  & 4.45 (8) & 10.0$^{+ 0.720}_{- 0.670}$ &  3.08$^{+ 0.190}_{- 0.140}$ & 15.5$^{+ 0.080}_{- 0.130}$ & 10.9$\pm$3.75 \\
P12 & 0.99  &  7.84 (19) & 8.30  (12) &  9.37  (6)  & 4.55 (6) & 10.0$^{+ 0.720}_{- 0.450}$ &  3.12$^{+ 0.140}_{- 0.110}$ & 15.4$^{+ 0.070}_{- 0.080}$ & 13.1$\pm$4.56 \\
P13 & 0.99  &  7.83 (18) & 8.06 (10) &  9.30  (6)  & 4.60 (5) &  9.12$^{+ 0.430}_{- 0.410}$ &  3.23$^{+ 0.120}_{- 0.100}$ & 15.3$^{+ 0.060}_{- 0.070}$  & 8.89$\pm$3.60 \\
P14 & 0.88  & 9.25  (32) & 9.12  (30) &  9.87  (10)  & 4.71 (4) & 10.5$^{+ 1.28}_{- 0.470}$ &  3.06$^{+ 0.280}_{- 0.190}$ & 15.5$^{+ 0.130}_{- 0.160}$  & 7.72$\pm$3.76 \\
P15 & 0.88  &  8.99  (37) & 9.01  (20) &  10.2 (1)  & 4.67 (7) & 10.5$^{+ 1.28}_{- 0.920}$ &  3.09$^{+ 0.680}_{- 0.220}$ & 15.3$^{+ 0.130}_{- 0.520}$  & 6.92$\pm$3.63  \\
P16 & 0.79  & 8.87  (13) & 8.80  (6) &  9.18  (12)  & 4.82 (10) & 10.5$^{+ 0.750}_{- 0.470}$ &  3.30$^{+ 0.090}_{- 0.090}$ & 14.9$^{+ 0.040}_{- 0.060}$  & 5.80$\pm$3.78  \\
P17 & 0.78  &  9.16  (12) & 9.23  (7) &  9.12  (11)  & 4.75 (8) & 10.5$^{+ 0.490}_{- 0.240}$ &  3.20$^{+ 0.070}_{- 0.070}$ & 15.1$^{+ 0.040}_{- 0.050}$ & 7.03$\pm$3.82  \\
\enddata
\end{deluxetable*}

\begin{figure}
\centering
\includegraphics[width=0.48\textwidth]{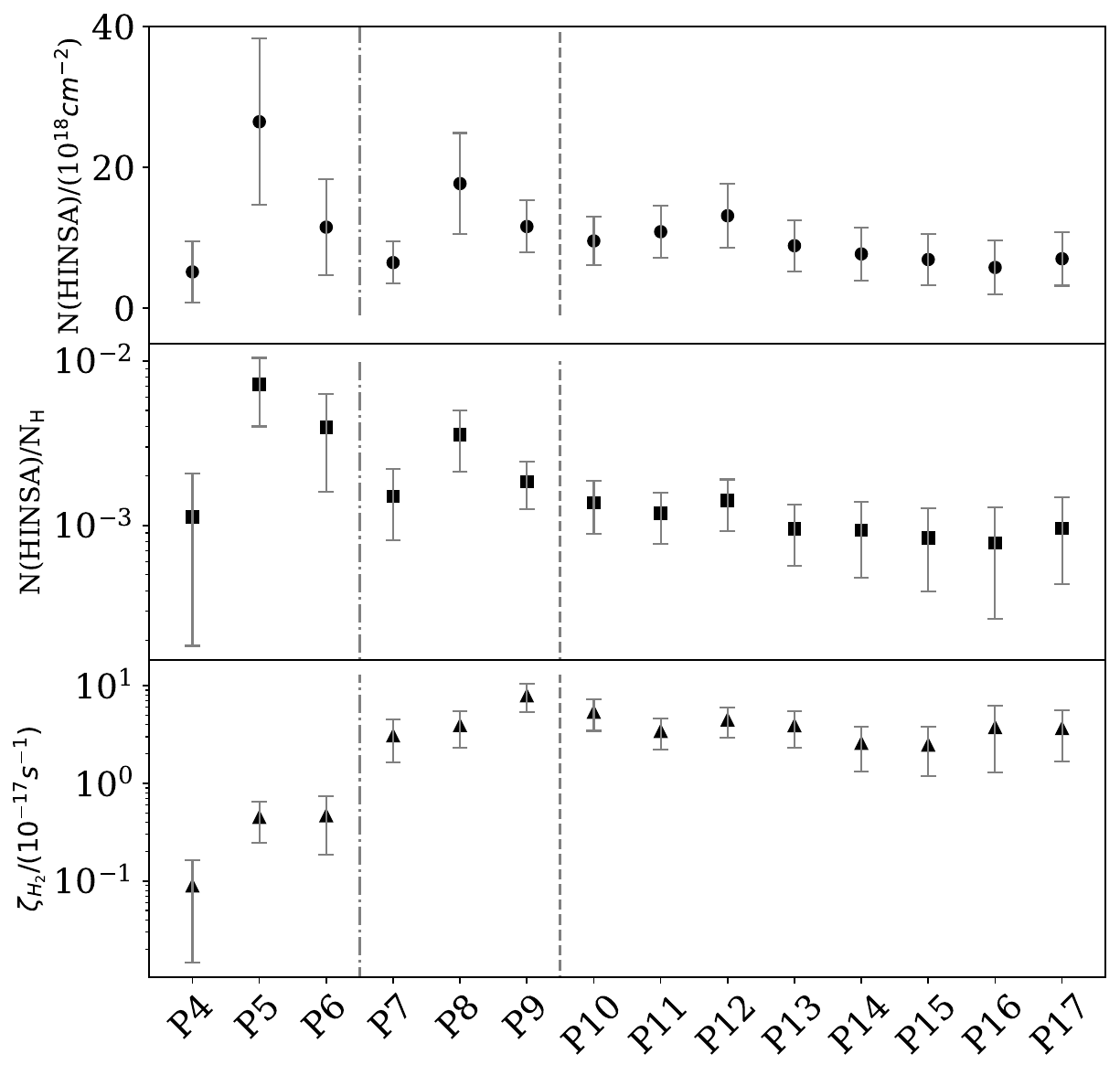}
\caption{HINSA column density N(HINSA) (top panel), HINSA abundance (middle panel) and  $\zeta\rm_{H_2}$ (lower panel)  toward P4 to P17. The vertical dash-dotted and dashed lines represent the spatial locations of  the \h2\ peak and CO edge shown in Fig. \ref{fig:boundpos}.}
\label{fig:hinsa}
\end{figure}

The total column density, $N_H$, is determined from the E(B-V) value with $N_H/E(B-V)=9.4\times 10^{21}$ \cm2 mag$^{-1}$ \citep{2018ApJ...862...49N}, not the canonical value of $5.8\times 10^{21} E(B-V)$ \cm2 mag$^{-1}$ \citep{1978ApJ...224..132B},  but one favored by recent studies \citep{2017ApJ...846...38L, 2019ApJ...886..108F}. The HINSA abundance is calculated as $x \rm =N(HINSA)/N_H$. As shown in Fig. \ref{fig:hinsa}, the HINSA abundance has a peak at  $5.9\times 10^{-3}$ toward P5 and  minimum value of $7.0 \times 10^{-4}$ toward P16.  These values are consistent with the value in the Taurus region, for instance,  $2.5\times 10^{-4}$ toward the L1544 cloud and $\sim 2\times 10^{-3}$  toward the L1574 cloud \citep{2003ApJ...585..823L}.

\section{Discussion}
\label{sec:discussion}

\subsection{Formation of Cold Dense Flow by Shock Compression}
\label{subsec:cold_flow}

Many numerical simulations have investigated the transition from \hi\ to \h2\ gas under various environments \citep[e.g.,][]{2010MNRAS.404....2G, 2010ApJ...715.1302V, 2012MNRAS.424.2599C, 2016A&A...587A..76V, 2022MNRAS.512.4765S}. In these simulations, the presence of shocks is considered to accelerate this process  \citep[e.g.,][]{2002ApJ...564L..97K, 2005ApJ...626..864M}.  Follow up simulations in realistic environment, e.g., collision of gas streams at moderately supersonic velocities  \citep{2006ApJ...643..245V, 2007ApJ...657..870V} and collision of superbubbles \citep{2011ApJ...731...13N}  found the formation of cold, dense, and filamentary structure in the collision zone. 

\citet{2015ApJS..219...20L}  identified 37 bubbles in the Taurus molecular cloud, suggesting the presence of shocks in this region. According to simulations, the presence of shocks can result in a sharp density increase of \h2\ across the shock front  \citep[e.g.,][]{1998MNRAS.295..672C}. As shown in Section \ref{sec:analysis}, an apparent jump  of \h2\ density and the appearance of cold \hi\ gas mixed with molecular gas are observed from the outer region into the inner region.  Our observations are consistent with simulations of shock-accelerated \h2\ formation.  However, further observations of shock tracers (e..g., SiO ($J=2-1$)) across the linear edge are essential to validate the proposed shock-induced \h2\ formation. 

 \subsection{The Relationship between Kinetic Temperature and Excitation Temperature}
 \label{subsec:tk_tex21}

With UV absorptions  of CO toward stars, the excitation temperature $T\rm_{ex}$  of  low-$J$ rotational lines of CO has been determined to be less than 5 K  in diffuse molecular  clouds \citep{2008ApJ...687.1075S}. These findings can be explained by assuming $T\rm_k$ values of 50 or 100 K \citep{2013ApJ...774..134G}. The notable disparity  between $T\rm_{ex}$ and  $T\rm_k$  suggests  subthermal excitation, indicating a departure from LTE conditions. 
 
As outlined in Table \ref{table:fitpara}, the $T\rm_{ex}$ values rise from approximately $ 5$ K to around $\sim 10$ K for multi-transitions of CO. However, the $T\rm_{ex}$ values derived from optically thin \13co\ transitions are notably lower, ranging from 2.80 to 4.82 K.  While both sets of calculations yield lower excitation temperature than the  $T\rm_k$ values of $10-39$ K obtained from the MCMC fit,  the discrepancy for optically thick lines (low-$J$ transitions of CO) is less pronounced. Calculation of physical parameters  under LTE conditions are likely to introduce less uncertainty for optically thick lines.

\subsection{Cloud Geometry Revealed by Multi-J CO Transitions}
 \label{subsec:cloud_geo}

\begin{figure}
\centering
\includegraphics[width=0.5\textwidth]{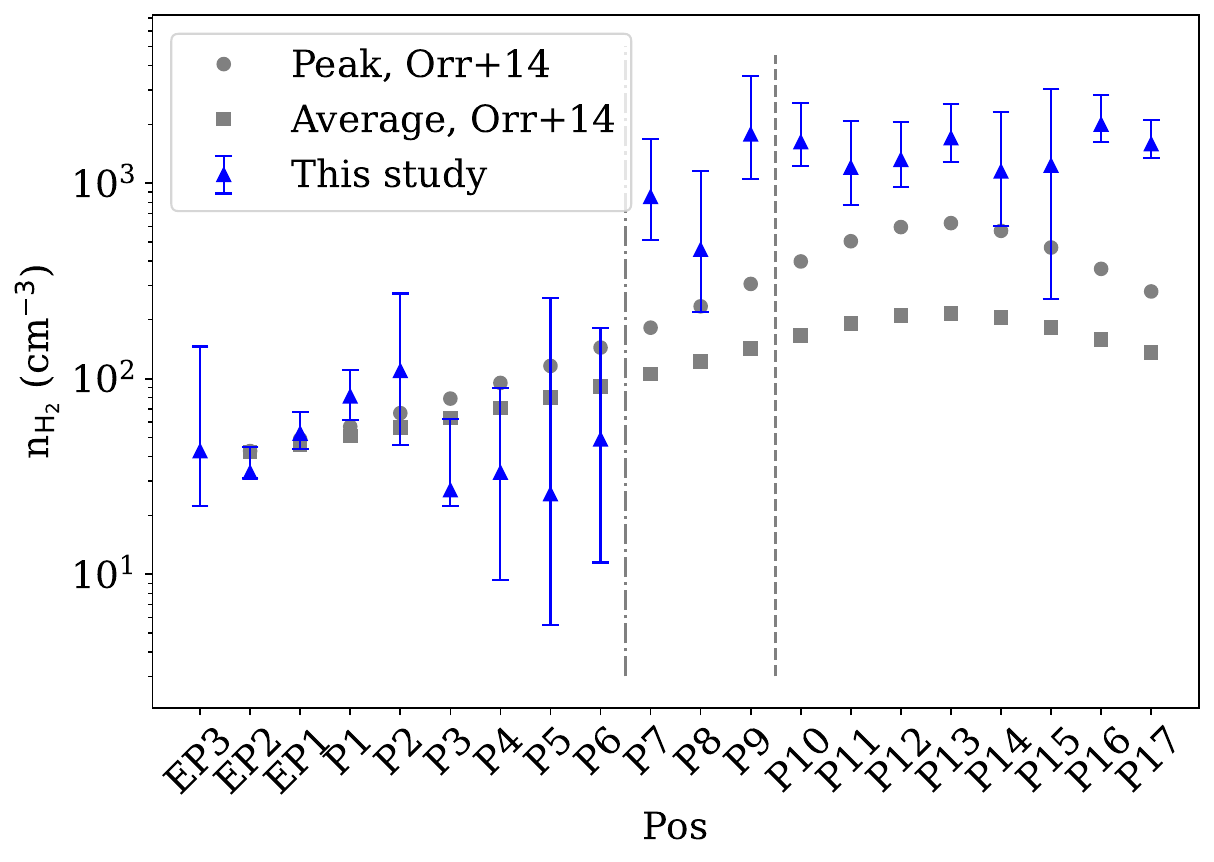}
\caption{Volume density of \h2, n$_{H_{2}}$ towards 20 positions. Peak and average density along line of sight are marked with black dots and blue triangles, respectively. The position of EP3 exceeds the radius of cylinder in Orr et al.\ (2014). Its density was adopted as 0 \cc.  The vertical dash-dotted and dashed lines represent the spatial locations of  the \h2\ peak and CO edge shown in Fig. \ref{fig:boundpos}.}
\label{fig:den_orr}
\end{figure}

With observations of CO, [C {\sc i}], and [C {\sc ii}],  \citet{2010ApJ...715.1370G} and \citet{2014ApJ...795...26O} proposed a cylindrical geometry to model this linear edge region with a two-dimensional RATRAN radiative transfer code. The fitted volume density  follows a power law with the distance to the core within a truncating radius, that is described by:

\begin{equation}
    n_{H_2}(r)= \left\{ \begin{array}{ccr}
         n_ca^2/(r^2+a^2) & \mbox{,  } & 
         r \leq R \\ 0  & \mbox{,  } & r > R                 \end{array},\right.
\end{equation}
where n$_{H_{2}}$ is the volume density of H$_{2}$, $r$ is the distance to the center of the cylindrical, n$_{c}$ is the maximum volume density corresponding to the center of the cylindrical, $a$ is the width of the central core, and R is the truncating radius. The value of n$_{c}$, $a$ and R are 626 cm$^{-3}$, 0.457 pc, and 1.80 pc, respectively.  With this model, the peak H$_{2}$ volume density towards 20 sightlines ranges from 50 to 620 \cc\ (Fig. \ref{fig:den_orr}) except EP3, whose position exceeds the radius in  \citet{2014ApJ...795...26O}.  

As shown in Fig. \ref{fig:den_orr},   the volume densities derived using the one-dimensional RADEX code in this study are significantly higher than the peak volume densities in the model of   \citet{2014ApJ...795...26O}, except for positions P3 to P6.  However, we note that our analysis incorporates higher-J transitions of \co\ and \13co (up to J=3-2), while excluding [C {\sc i}] and [C {\sc ii}] lines, which primarily trace the diffuse gas component. This methodological difference likely explains the higher volume densities derived in our model  than that in \citet{2014ApJ...795...26O}.

Beside the inconsistency of volume density, the cylindrical geometry can not explain the velocity difference of $1-3$ \kms\ between two velocity components, which is clearly seen in CO(1-0) and CO(2-1) data. The presence of two velocity components is observed in OH emission too \citep{2016ApJ...819...22X}. There exists a U shape around the linear edge (see details of Fig. \ref{fig:boundpos}), which is consistent with the prediction of numerical cloud-cloud simulation \citep[e.g.,][]{2024JPhCS2796a2005N}. The consistency suggests that this linear edge may be due to the collision of a massive gas stream (the first velocity component centering $5.6-6.7$ \kms) and a gas stream  with a much lower \h2\ column density(the second velocity component centering $7.7-8.6$ \kms).

\subsection{Low-energy Cosmic Ray Ionization Rate across the Linear edge}
\label{subsec:hi_h2}

The cold \hi\ responsible for the HINSA feature originates from the dissociation of \h2\ by low-energy cosmic rays. When the dissociation of \h2\ reaches equilibrium with the formation of \h2\ on dust grains, the abundance of cold \hi\ remains constant over time, indicating a steady-state  condition. Under this condition, the low-energy cosmic ray ionization rate of \h2\,  $\zeta\rm_{H_2}$ can be expressed with the formulation in \citet{2005ApJ...622..938G},
\begin{equation}
\zeta_{H_2} = \frac{2k'n_0 x}{1- x},
\label{eq:zeta_h2}
\end{equation}
where   $k'$, $n_0$ and $x$ represent  the formation rate coefficient of \h2\ , the total volume density,  and the \hi\ abundance, respectively. The composition and size distribution of grains can strongly affect the value of $k'$ by an order of magnitude \citep{2004A&A...414..531H, 2005ApJ...622..938G}. We have adopted the general expression from \citet{2005ApJ...622..938G}, which gives $k' = 1.2 \times 10^{-17}\ \text{cm}^3\ \text{s}^{-1}$ at $T\rm_ k= 10$ K.  

As shown in Fig. \ref{fig:hinsa}, the value of $\zeta\rm_{H_2}$ peaks at a maximum of ($7.9\pm 2.5$) $\times 10^{-17} s^{-1}$ toward P9. This closely aligns with the range of ($3-6$) $\times 10^{-17}$ s$^{-1}$ in \citet{2014ApJ...795...26O}. However, the ionization rate experiences a  two-order-of-magnitude variation, reaching approximately  $(8.9 \pm 7.5) \times 10^{-19}$ s$^{-1}$ toward P4. This variability can be attributed to the significant changes in number density observed across the linear edge. 

\subsection{Influence and Physical State of the Second Velocity Component}
\label{subsec:second_comp}

The presence of the second velocity component is notably evident in the CO(1-0) and CO(2-1) emissions spanning from P14 to P17. Determining the physical properties of the second velocity component would significantly benefit unraveling its origin. However, this task is challenging due to the lack of observations of optically thin lines. To enhance constraints on the physical environment, we assume a beam filling factor of 1 for the second velocity component. With this assumption, we simultaneously fit the intensity of CO(1-0) and CO(2-1) by varying $n\rm_{H_2}$, $T\rm_k$ and the \co\ column density $N$(CO). 

The fitting results suggest that the second component has an \h2\ density of $n\rm_{H_2}\sim 1500$ \cc, a kinetic temperature of $T\rm_k \sim 15$ K, and a CO column density of $N$(CO) $\sim 3\times 10^{15}$ \cm2 .  Assuming the same CO abundance of  $3\times 10^{-5}$ toward P10, the \h2\ column density of the second component  is estimated to be approximately $1 \times 10^{20}$ \cm2. These results  suggest that the second component is a dense molecular cloud with a relatively low N(\h2) value.   Further observations are needed to better characterize its physical properties.

\section{Summary}
\label{sec:summary}

We  derived multiple transitions of CO and \13co\ data toward 20 positions across the linear edge of the Taurus molecular cloud. By combining the CO data with archival \hi\ data, we  obtained physical properties and HINSA abundance across the edge.   Quantitative comparison with hydrodynamical simulations tentatively indicate that colliding gas flow shocks could drive the atomic-to-molecular phase transition along the linear edge. The main results can be summarized as follows: 

\begin{enumerate}
\item  Two velocity components with varying central velocities are identified. The main velocity component, centered at $\sim 6$ \kms\, is detected in all multi-$J$ CO data  while the second component, centered at $\sim 8$ \kms\, is observed in the CO(1-0) and CO(2-1) data within the linear edge.  

\item  The intensity ratio between \co\ and \13co\ in the $J=2-1$ transition implies a lower limit for the  $^{12}C/^{13}C$ ratio of $54\pm 17$. This value is consistent with those found  in the local ISM and with the value of 90 inferred for TMC-1  from recent HCCNC and HNCCC observations.  

\item  The excitation temperature ranges from 5 to 10 K when derived from  CO(1-0) data under the optically thick assumption. Assuming  optically thin conditions and LTE, the excitation temperature varies from 2.80 to 4.82 K based on the intensity ratio between  \13co(2-1) and \13co(1-0) data. These are significantly lower values compared to  typical kinetic temperatures in molecular clouds,  indicating  subthermal excitation conditions. 

\item  We combine MCMC and non-LTE RADEX analyses to calculate the physical properties at these positions. Under a one-dimensional uniform geometry, kinetic temperature is found to range from 8.71 K to 436 K. Additionally, the number density ranges from 7.0 to $2.0\times 10^3$ \cc,  while N(\13co) varies from $1.20 \times 10^{14}$ to $3.02 \times 10^{15}$ \cm2 (Fig. \ref{fig:mcmc_paras}).  An obvious jump of number density appears in the position with peak \h2\ intensity, which does not align with the cylindrical model proposed by   \citet{2014ApJ...795...26O} (Fig. \ref{fig:den_orr}).

\item Cold \hi\ gas appearing as HINSA feature is detected across the linear edge.  Assuming a  steady state,  the  HINSA abundance indicates a peak low-energy cosmic ray ionization rate of $\sim 6.9 \times 10^{-17} s^{-1}$ toward P9 (Fig. \ref{fig:hinsa}), the position outside the \13co(1-0) edge.  

\item The second velocity component can be characterized with physical properties of $n\rm_{H_2}\sim 1500$ \cc, $T\rm_k \sim 15$ K and $N$(CO) $\sim 3\times 10^{15}$ \cm2. This component exhibits a similar number density and kinetic temperature but a smaller column density compared to the main velocity component.  One possible explanation is that the linear edge could be a result of the collision between the main and the secondary velocity components.

\end{enumerate}

\section*{Acknowledgments}

We are grateful to the anonymous referee for the constructive suggestions, which have greatly improved this paper. 
We appreciate Zhiyu Zhang and Jing Zhou for their valuable suggestions regarding the use of RADEX and MCMC. This work is  is sponsored by the National Natural Science Foundation of China (grant No. 11988101, 12473023), The University Annual Scientific Research Plan of Anhui Province (No. 2023AH030052, No. 2022AH010013), the National SKA Program of China (grant No. 2022SKA0120101), the China Manned Space Program through its Space Application System, Zhejiang Lab Open Research Project (NO. K2022PE0AB01), Cultivation Project for FAST Scientific Payoff and Research Achievement of CAMS-CAS. Di Li is a New Cornerstone Investigator and is supported by National Key R\&D Program of China No. 2023YFE0110500, grant NSF PHY-2309135 to the Kavli Institute for Theoretical Physics (KITP).

The  University of Arizona  Submillimeter Telescope on Mt. Graham is operated by the Arizona Radio Observatory (ARO), Steward Observatory, University of Arizona. Part of CO data were observed with the Delingha 13.7 m telescope of the Qinghai Station of Purple Mountain Observatory. We appreciate all the staff members of the Delingha observatory for their help during the observations.


\end{CJK}
\end{document}